\documentstyle[twocolumn,aps,prb,epsf]{revtex}
\begin{document}
\preprint{NUSc/00-05}
\title{Flux Phase as a Dynamic Jahn-Teller Phase: \\
Berryonic Matter in the Cuprates?}

\author{R.S. Markiewicz and C. Kusko}

\address{Physics Department and Barnett Institute, 
Northeastern U.,
Boston MA 02115}
\maketitle

\begin{abstract}
There is considerable evidence for some form of charge ordering on the
hole-doped stripes in the cuprates, mainly associated with the low-temperature 
tetragonal phase, but with some evidence for 
either charge density waves or a flux phase, which is a form of dynamic 
charge-density wave.  These three states form a pseudospin triplet,
demonstrating a close connection with the $E\otimes e$ dynamic Jahn-Teller 
effect, suggesting that the cuprates constitute a form of Berryonic matter.  
This in 
turn suggests a  new model for the dynamic Jahn-Teller effect as a form of flux 
phase.  A simple model of the Cu-O bond stretching phonons allows an estimate of
electron-phonon coupling for these modes, explaining why the half breathing 
mode softens so much more than the full oxygen breathing mode.  The anomalous
properties of $O^{2-}$ provide a coupling (correlated hopping) which acts to
stabilize density wave phases.
\end{abstract}


%****
\narrowtext
%****

\section{Introduction}

In a (molecular) Jahn-Teller (JT) effect, an electronic degeneracy in a 
symmetric molecule is lifted by distorting the molecule and lowering the free
energy by an amount $E_{JT}$.  A {\it dynamic} JT (dJT) effect arises when there
is a degeneracy both of electronic levels and of molecular vibrations (strictly 
speaking, when two independent distortions have the same JT energy lowering).  
Quantum tunneling between the two distortions restores the (time-averaged)
symmetry of the molecule, but leads to a net orbital angular momentum which is
half-integrally quantized\cite{HLH,Del}.  In principle, an analog of the dJT 
effect should exist in dense solids\cite{EnHa,JT1}, but this feature of ground 
state nuclear motion has not been adequately described.  Recently, a number of
proposals have been made for solid state systems with finite orbital angular 
momentum in their groundstates, including `Berryonic matter'\cite{Berr} and
the flux phase\cite{Affl,Laugh,flux2,Capp} or orbital 
antiferromagnet\cite{HaRi,Sch}.  Here the connection between these concepts is
worked out, a formalism is presented for describing the dJT effect in terms of 
phonon-assisted flux phases.

This result is particularly timely, since dJT effects have been proposed to play
a role in a number of materials of current interest, including
buckyballs\cite{Auer,Berr}, cuprates\cite{KAM,RM8b,RM8cd}, and 
manganites\cite{AJM}.  Remarkably, many of these materials display stripe phases
and/or high temperature superconductivity, and it has been proposed that either 
strong electron-phonon coupling\cite{RM3,ZXph} or flux phases may be responsible
for these features.  Moreover, in the cuprates, there is recent evidence for 
both the flux phase\cite{Moofl} and for strong phonon 
anomalies\cite{ZXph,McQ1,McQ2,McQ4,McQ3,McQx,Bian}.  Hence, the present 
formalism is applied to a simple model of the in-plane phonons in the cuprates.

The two JT modes plus the dJT state can be combined into a `pseudospin'
triplet\cite{BAPS} of site charge-density wave (CDW), bond CDW, and flux phase 
(dynamical CDW).  Restricted to a single four-copper plaquette `molecule', they 
reduce to an $E\otimes (b_1+b_2)$ JT problem.  The site and bond CDW's can 
be related to important phonon modes in the cuprates, namely the oxygen 
breathing mode and a shear mode which couples to the tilting mode associated 
with the low-temperature tetragonal (LTT) phase.  Since there is considerable 
interest in the flux phase, some care is taken to analyze it in terms of the 
associated structural distortions.  In particular, the anomalous $\pi$ phase is 
just the Berry phase of the dJT effect.  In the molecular problem, the
dJT effect arises at a (weak) degeneracy point; remarkably, at this point
the dynamics is (classically) chaotic.  

Our model calculations find a crossover from an antiferromagnetic insulator at
half filling to a doped paramagnetic phase stabilized by density wave order.
(The detailed doping dependence is probably complicated by the presence of
stripes, as discussed in the Appendix.)  The density wave is predominantly
stabilized by electron-electron interaction, but with a significant structural
component.  The system is close to but probably not at the dJT degeneracy
point, with the shear and flux phases lying lowest in energy.

This paper is organized as follows.  Section II surveys the various phonons 
which experimentally are found to couple to the holes, and shows that they fall 
into three classes, including site and bond CDW's and the flux phase.  It is 
explained how the bond CDW couples to LTT and dimpling distortions.  
(In an Appendix additional evidence is presented that the coupling
is particularly strong in the hole-doped regions, i.e., on charged stripes.)  
Section III shows that these classes have a well-defined group 
(pseudospin) structure (Section IIIA), and are closely associated with the 
dJT effect of a square molecule (IIIB).  
Section IIIC generalizes the pseudospin formalism to a lattice.  In Section IV 
these results are applied to the cuprates, using both a three band (IVA) 
and a one band (IVB) model to estimate the electron-phonon coupling of a large
number of planar bond stretching modes, including some outside of the above 
three classes.  The reason why the half breathing mode is so strongly 
renormalized is explained.  A simple model of phononic flux phases is presented 
which can provide a basis for a general theory of band dJT effects.  
In Section V the self-consistent gap equations are solved for combined
electron-electron and electron-phonon coupling, and it is found 
that the anomalous behavior of the $O^{2-}$ molecule leads to a particular form 
of correlated hopping which helps stabilize paramagnetic CDW phases.  
Some discussion of {\it desiderata} for a more complete model are presented in
Section VI, and conclusions are presented in Section VII.

\section{Classification of Phonon Modes}

There is considerable evidence for strong electron-phonon coupling effects in
the cuprates.  While the effects are weaker than in the nickelates and
manganites, we postulate that it is this relative weakness which allows
superconductivity to compete against charge ordering.  The difficulty in 
analyzing the coupling lies in the fact that it is spread over so many different
modes, and the modes vary from cuprate to cuprate.  Here we show that the
strongly coupling modes can be classified into three groups: site and bond
CDW's and flux phases.  Additional evidence specifically tying the phonons to
the charged stripes is presented in the Appendix.

The three categories of phonons can be described in different ways.  Perhaps the
simplest is as site CDW (leading to pileup of charge on alternate Cu's) vs bond
CDW's (pileup on oxygens) vs dynamic interchange of the two (the flux or orbital
antiferromagnetic -- OAF -- phase).  Alternatively, since the modes all couple
strongly to a Van Hove singularity (VHS), the first two may be 
classified by how they split the VHS peak: the site CDW's leave the $X$ and
$Y$ (($\pi ,0$) and $(0,\pi )$) point VHS's equivalent, but split each (as in 
the one-dimensional Peierls transition); the bond CDW's make the $X$ and $Y$ 
points inequivalent.

\subsection{Breathing Modes}

Strong coupling to oxygen breathing modes is found in a number of cuprates.  
Weber\cite{Web} originally suggested that there should be large coupling of the 
electrons to the oxygen breathing mode at $(\pi ,\pi )$.  However, this mode is 
suppressed near half filling by strong coupling effects, since the CDW 
requires a charge imbalance on different copper atoms.  The coupling is 
enhanced by proximity to the VHS, and hence could lead to a CDW instability near
optimal doping\cite{RM8b,RM3,Cast,Surv}.  Experimentally, 
breathing modes appear more strongly coupled near $(\pi ,0)$ 
than at $(\pi ,\pi )$, both in La$_{2-x}$Sr$_x$CuO$_4$ (LSCO) and 
YBa$_2$Cu$_3$O$_{7-\delta}$ (YBCO)\cite{PintR}.  McQueeney, et 
al.\cite{McQ1} have found that the phonon spectra show evidence for a dynamic 
period doubling in the $\Gamma -(\pi ,0)$ direction, which they associate with 
stripes.  Since this period is independent of doping\cite{McQ2}, it is unlikely 
to be related to stripe periodicity, but could be associated with charge 
ordering {\it along} the stripes.  A possible explanation for the strong 
coupling of
this half breathing mode is given in Section IVA.  Very similar anomalies are 
found in the nickelates, with phonon softening and splitting over a large part 
of the Brillouin zone, including evidence for local modes tied to 
stripes\cite{TraNO}.  In LSCO, stripe correlations have a strong effect on 
these oxygen-related phonons near 70 meV\cite{McQ4,McQ3}.

In this same energy range (50-70meV below the Fermi level), photoemission 
measurements\cite{PEkink,PEk2} have found a break in the dispersion of the 
cuprates, present throughout $k$-space, although having much more dramatic 
effects near $(\pi ,0)$.  While there are alternative 
interpretations\cite{PEk2}, it has been proposed\cite{ZXph} that this is the 
conventional renormalization associated with strong electron-phonon 
coupling\cite{AM} to the breathing mode phonons near this energy.  Such 
renormalization can produce the extended saddle points observed near 
$(\pi ,0)$\cite{RMXB}.  

\subsection{Relation of Shear Mode to LTT, Dimpling, etc.}

Evidence for proximity to a shear mode instability has been noted in LSCO near 
the 1/8th anomaly, in the form of softening of the $C_{11}-C_{12}$ elastic 
modulus\cite{SSNN}.  However, it is remarkable that, whereas theoretically 
there should be a strong electron-phonon coupling with {\it bond stretching} 
modes, the observed soft modes are {\it bond bending} modes, with different 
modes in each family of cuprates: the low-temperature orthorhombic (LTO) and 
LTT tilts in LSCO, a dimpling mode in YBCO, and a rotation of the four planar 
oxygens about the central Cu in the electron-doped cuprates\cite{Brad33}.
Here, we suggest that there is a connection between these modes and the shear
instability.

We first consider LSCO.  
It is a common finding that the local structure in the cuprates differs 
from the average structural order determined by neutron or x-ray 
diffraction\cite{EgGins}, and it has been suggested that the observed LTO phase
in LSCO is really dynamically disordered, with the local order being closer to
LTT\cite{RM8cd}.  There have been numerous hints of local LTT 
order\cite{Bia2,LTT1,LTT4,LTT2,LTT3,Boz}, but the problem is complicated by the
simultaneous occurence of stripe order.  The issue of the relation between
stripe order and the LTT distortion is addressed in the Appendix.
Here, we merely note that the relevant local distortion is likely to be of LTT
form, and explore the connection of this distortion to the shear mode.

By increasing the Cu-O bond length along the tilted bonds, both the shear and 
the LTT distortions make the two oxygens in a unit cell crystallographically 
inequivalent, thereby coupling to the VHS.  Such VHS splitting has recently 
received renewed interest\cite{HaM,VV}.  We have suggested that the LTT tilt 
is a secondary modification of a shear mode instability due to lattice mismatch 
effects\cite{RMXB}: the underlying instability is to a shear (making
inequivalent bondlengths), but the CuO$_2$ planes are under compression, so the
LTT tilt arises as a compromise, making the Cu-O bondlengths inequivalent,
without increasing the net Cu-Cu bondlength.  Qualitatively, the LTT tilt 
pattern arises because the CuO$_6$ octahedra tilt as rigid units\cite{RUM}: 
since the octahedra are corner sharing, alternate octahedra must tilt
in opposite directions, leading to a distortion at wave vector $\vec Q=(\pi ,
\pi )$.
(The detailed interpretation involves partial hole localization on oxygens, and
the large size difference between $O^-$ and $O^{2-}$.\cite{RMXB})

Similar effects arise in other cuprates.  
Thus, high energy phonons behave similarly in LSCO and YBCO, but at low 
frequencies there are very different structural distortion. In YBCO 
there is little evidence for an LTT phase (although there was a report of a
local tilt-mode instability\cite{Schw}).  More likely, the role of the LTT
distortion is taken by the {\it dimpling mode}.  While this mode involves 
predominantly copper and oxygen motion out of the CuO$_2$ planes, it also 
couples to in-plane Cu-O displacements.  In particular, it couples to the
orthorhombicity\cite{Roh}, and hence to the same shear mode involved in the LTT
phase (distortion $B_1$ in Fig.~\ref{fig:2} below).  Hence the dimpling also 
splits the VHS's, but in a different pattern\cite{OKA}.  Experimentally, the
distortion competes with superconductivity, and a large decrease of the 
orthorhombicity is found both at and below $T_c$\cite{VP}, while a discontinuous
change in the dimpling is found on passing through optimal doping\cite{Kal}.  
{\it Chain ordering} also plays a role, since it splits the degeneracy of the 
VHS's along and across the chain direction.  Photoemission seems to find that 
the VHS along X (Y) is below (above) the Fermi level\cite{Schnab}, while neutron
scattering\cite{Mook} finds that the stripes align along the chain direction, 
reminiscent of LTT pinning.  In a related 123 compound, 
(La$_{1-x}$Ca$_x$)\-(Ba$_{1.75-x}$La$_{0.25+x}$)Cu$_3$O$_y$, the La$^{3+}$ at 
the Ba cite disrupts the chains, leading to an average tetragonal structure;
in this case a plane buckling is observed, which is {\it largest at the doping
of optimal superconducting $T_c$}.\cite{Chm}

A similar effect is found in Bi$_2$Sr$_2$CaCu$_2$O$_{8+\delta}$ (Bi2212).  The 
`ghost' Fermi surfaces in Bi2212 are clear evidence for a lattice period 
doubling, and a connection with the antiferromagnetic order has been 
suggested\cite{Aeb}.  However, since they are present at room temperature and at
optimal doping -- far from the region where strong magnetic correlations would 
be expected -- they may instead be related to a structural transition -- the 
known orthorhombic ordering\cite{SiPi}.  This lattice anomaly also 
couples strongly to the VHS.  Most suggestively, it has recently been found that
{\it the intensity of these ghost features peaks near optimal 
doping}\cite{KoBo}.  This is close to the prediction of the VHS stripe model, in
which the structural order should be maximal in the slightly overdoped regime, 
just when magnetic correlations disappear (termination of the stripe phase).

\subsection{The Flux Phase}

The flux phase has generated intense interest because of its highly unusual 
physical properties.  The experimental evidence for this phase, which remains 
controversial, includes (i) the fit of the dispersion of the Mott insulator 
Sr$_2$CuO$_2$Cl$_2$ (SCOC)\cite{Well} to the flux phase form\cite{Laugh} 
(however, alternative interpretations include a Mott antiferromagnet, while more
recent experiments find a more complex dispersion\cite{SCOC}), (ii) the 
d-wave-like nature of the normal-state pseudogap\cite{Mes}, which has been 
interpreted in terms of the conical point (node) in the flux phase dispersion at
$(\pi /2,\pi /2)$, and (iii) the recent observation of the induced magnetic 
fields associated with the orbital currents\cite{Moofl}.  (However, more recent
measurements\cite{Moof2} find that the magnetic moment lies in-plane, and not 
along the c-axis as originally expected, and the experiment also has alternate 
interpretations\cite{Allo}.)

\subsection{Competition and Cooperation of the CDW's}

In Section IIIB, we will show that the molecular equivalent of the flux phase,
the dJT state, arises from competition between the two other modes.  In
the cuprates, there is also evidence for a competition between the two types of
CDW, which we briefly review here.  In LSCO, both the breathing and the LTT 
modes show anomalous softening with doping.  An interesting situation arises in 
YBCO.  Pintschovius and Reichardt (p. 349 of Ref.~\onlinecite{PintR}) note that 
hole doping may renormalize the half-breathing mode phonon frequency from 74meV 
to $\sim$37meV, due to a special form of Van Hove nesting\cite{OKA}.  The 
mechanism proposed involves VHS induced coupling of two phonon modes 
remarkably similar to the flux phase effects we are describing.  The dimpling
of the CuO$_2$ planes causes the VHS's to bifurcate (e.g., the $X$ VHS 
splits along the $\Gamma -X-\Gamma$ line).  In turn, nesting between the 
bifurcated peaks leads to breathing mode softening along $\Gamma -X$ and can
lead to a period doubling instability.  The exact nesting wave vector is 
controlled by the degree of VHS bifurcation, which in turn depends on the amount
of dimpling.  Thus, two phonon modes (here the dimpling and half-breathing 
modes) are strongly coupled by the VHS.  A similar competition between breathing
and octahedral tilt modes near a superconducting instability is 
found\cite{Bradx} in Ba$_{1-x}$K$_x$BiO$_3$; in this case the tilt modes are 
near the $R$ point of the Brillouin zone, coupling to the three-dimensional 
VHS's.

Other routes to mode coupling are also possible.  For example, when the 
breathing mode softens\cite{PintR,McQ1,McQ2}, it can couple to a lower frequency
phonon, either an LTT distortion in LSCO or a dimpling mode in YBCO, and drive 
that second mode soft.  This strong mixing of the two modes can create a flux 
phase. 

\section{Electron-Phonon Coupling: From Square Molecules to Cuprates}

In this section, we show that the three phonon modes form a symmetry group, and
map the interaction to a molecular JT model.  We demonstrate that the same
degeneracy arises in both the molecular and lattice problems.  In Section IV we
apply these results to the cuprates, showing how the phonons in LSCO fit into 
the scheme, and calculating their coupling to electrons.

\subsection{The Pseudospin Group}

We begin with a purely theoretical description, based on the group theory of
instability, or spectrum generating algebras\cite{SoBir}, specialized to the VHS
instability group\cite{Sch,MarV}, of SO(8) symmetry.  Within this group there is
a triplet of CDW operators, which couple strongly to phonons.  These operators 
are spanned by an $SU(2)$ subalgebra which we call `pseudospin'.  

To specify the order parameters in terms of bilinear electronic operators, it is
necessary to define these operators on {\it a larger unit cell}, containing a
plaquette of four copper atoms.  
It is convenient to introduce creation operators $a^{\dagger}_{\pm,\sigma}(\vec 
r) =(\psi^{\dagger}_{E_{u1},\sigma}(\vec r)\pm\psi^{\dagger}_{E_{u2},\sigma}
(\vec r))/2$.
The pseudospin can be written in terms of the Fourier transforms of these 
states.  In the basis 
\begin{equation}
\{B_{\sigma}(k)\}=\{a_{+,\sigma}(\vec k),a_{-,\sigma}(\vec k)\},
\label{eq:4}
\end{equation}
the explicit form of these matrices is
\begin{equation}
\tilde\tau ={\tilde\gamma\over 2}\tilde T_x,
\label{eq:12b}
\end{equation}
\begin{equation}
\tilde O_{OAF}={\tilde\gamma\over 2}\tilde T_y,
\label{eq:12c}
\end{equation}
\begin{equation}
\tilde O_{CDW}=\tilde T_z,
\label{eq:12a}
\end{equation}
where
\begin{equation}
\tilde T_x=\left(\matrix{0&1\cr
              1&0\cr}\right),
\label{eq:2b}
\end{equation}
\begin{equation}
\tilde T_y=\left(\matrix{0&-i\cr
              i&0\cr}\right),
\label{eq:2c}
\end{equation}
\begin{equation}
\tilde T_z=\left(\matrix{1&0\cr
            0&-1\cr}\right),
\label{eq:2a}
\end{equation}
with $c_i=\cos{k_ia}$ and $\tilde\gamma=c_x-c_y$.
The operators $\tilde O_{CDW}$, $\tilde\tau$, and $\tilde O_{OAF}$ transform 
exactly as a pseudospin near the VHS's ($c_x-c_y=\pm2$), but not at a general 
point in the Brillouin zone.  This same problem arises in SO(5)\cite{Zhang}, and
the Henley operator\cite{Hen} was introduced to deal with it.

\begin{figure}
\leavevmode
   \epsfxsize=0.33\textwidth\epsfbox{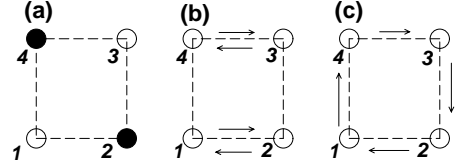}
\vskip0.5cm 
\caption{Pseudospin triplet: O$_{CDW}$ (a), $\tau$ (b), O$_{OAF}$ (c).}
\label{fig:1a}
\end{figure}
The electronic arrangements corresponding to these operators are illustrated in 
Fig.~\ref{fig:1a}: $O_{CDW}$ is a site CDW, with excess charge on the odd sites 
1 and 3, and a deficit on the even sites 2 and 4; $\tau$ is a bond centered CDW,
with excess charge on the 1-2 and 3-4 links, and a deficit on the 1-4 and 2-3 
bonds; while $O_{OAF}$ is an orbital antiferromagnet, with an orbital current 
flowing clockwise, from $1\rightarrow 4\rightarrow 3\rightarrow 2$. In a 
three-band model, $O_{CDW}$ would represent a Cu-based CDW, and $\tau$ an 
O-based CDW.  Note that the CDW mode is equivalent to the charge ordering found 
in La$_{2-x}$Sr$_x$NiO$_4$ for $x\sim 0.5$\cite{ZL}.  The orbital current 
operator\cite{Sch} $O_{OAF}$ is closely related to the flux phase\cite{Affl}.  
In the flux phase, hopping is accompanied by an accumulation of a phase change 
by $\pi$ on circulating a plaquette.  For the plaquette of Fig.~\ref{fig:1a}c, 
this leads to a contribution to the Hamiltonian proportional to
\begin{eqnarray}
H^{\prime}=e^{i\pi /4}(a^{\dagger}_2a_1+a^{\dagger}_3a_2+a^{\dagger}_4a_3+
a^{\dagger}_1a_4)+h.c.\nonumber \\
={1\over\sqrt{2}}(a_2^{\dagger}a_1+a^{\dagger}_3a_2+a^{\dagger}_4a_3+a^{\dagger}
_1a_4+h.c.)-2\sqrt{2}T_y.
\label{eq:11}
\end{eqnarray}
The first term in Eq.~\ref{eq:11} corresponds to an (uninteresting) uniform
hopping on the plaquette.  Thus, up to this term, {\it the flux phase is
equivalent to the orbital antiferromagnet $O_{OAF}$ ($T_y$)}.  

\begin{figure}
\leavevmode
   \epsfxsize=0.33\textwidth\epsfbox{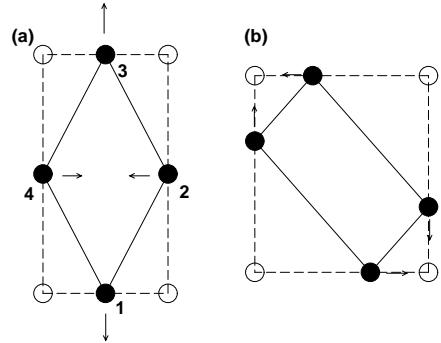}
\vskip0.5cm 
\caption{$B_1$ (a) and $B_2$ (b) distortions of Cu$_4$O$_4$.  Open (filled) 
circles represent the Cu's (O's).}
\label{fig:2}
\end{figure}
While the instabilities are predominantly electronic, they are accompanied by 
lattice instabilities, as in the Peierls transition.  In particular, the CDW 
couples to the oxygen breathing mode and the $\tau$-mode couples naturally to a 
shear distortion; these are referenced to a single molecule in Fig.~\ref{fig:2}.
The relation of this shear to the LTT tilting mode was clarified in Section II.

\begin{figure}
\leavevmode
   \epsfxsize=0.33\textwidth\epsfbox{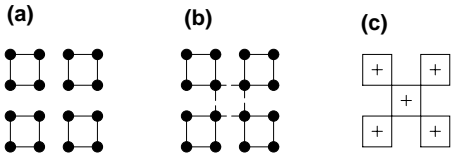}
\vskip0.5cm 
\caption{Berryonic matter: two views of the lattice of square molecules, either
as simple square (a) or as centered square with corner sharing molecules (b)
compared with the flux phase (c), where the $+$'s indicate plaquettes with a
$+\pi$ flux.}
\label{fig:41y}
\end{figure}

This pseudospin triplet is an exact condensed-matter analog to the molecular JT 
modes responsible for the dJT effects found in molecules.  The enlarged
unit cell associated with Eq.~\ref{eq:4} provides a natural connection. 
Thus, we regard the CuO$_2$ planes as composed of a square array of Cu$_4$O$_4$ 
`molecules' bonded to each neighbor by a pair of oxygens.  
These molecules can be embedded into the lattice in two equivalent
ways, as illustrated in Fig.~\ref{fig:41y}.  While it is simpler to analyze a
square lattice of square molecules, as in Fig.~\ref{fig:41y}a, a more accurate
representation is in terms of a `centered square' lattice as in 
Fig.~\ref{fig:41y}b.  [Crystallographically, the centered square is not a
primitive unit cell, but it is convenient for the present purposes.]  The latter
case has twice as many molecules, but they are corner sharing.  The atoms 
(coppers and oxygens) in each molecule deform in the same way.  For instance, 
Fig.~\ref{fig:41y}c illustrates the flux phase, labelling each square plaquette 
which contains a $+\pi$ flux with a plus sign.  When interpreted as in 
Fig.~\ref{fig:41y}b, the ground states have a two-fold degeneracy, which is 
also found in the flux and CDW phases.

The above analysis resolves an old conundrum.  Earlier calculations 
interpreted the LTT phase in terms of a Van Hove -- Jahn-Teller (VHJT) 
effect\cite{RM8cd,Surv}, where the electronic degeneracy is associated
with the presence of two VHS's.  This is puzzling, since JT effects usually
have a simple molecular basis, but such a basis is not obvious in the present 
case.  We now see that the VHJT effect on the conventional lattice is equivalent
to a conventional JT effect on a supercell lattice.  The resulting lattice 
constitutes a nearly exact realization of a Berryonic crystal -- replacing the 
originally proposed\cite{Berr} JT active triatomic molecules by square 
molecules.
\par
The close connection between the JT and Van Hove viewpoints can be clarified by
representing the CuO$_2$ plane electronic states in a supercell 
representation\cite{RM8cd,MarV}, replacing the copper (or Zhang-Rice singlet)
wavefunctions by electronic states symmetrized on a single plaquette.  These 
states $\psi_{+ijk}=(\psi_1+i
\psi_2+j\psi_3+k\psi_4)/2$ have $A_{1g}$ ($\psi_{++++}$), $B_{2g}$ ($\psi_{+-+-}
$), and $E_{u}$ ($\psi_{++--}$, $\psi_{+--+}$) symmetry.  Including nearest 
($t$) and next-nearest neighbor ($t'$) hopping, the kinetic energies are
\begin{eqnarray}
E(A_{1g})=-2t-t',
\nonumber \\
E(E_{u})=t',
\nonumber \\
E(B_{2g})=2t-t',
\label{eq:11z}
\end{eqnarray}
so at half filling, the $A_{1g}$ level is filled and there are two electrons in
the $E_{u}$ levels.  Optimal doping corresponds to one extra hole per plaquette,
leaving one electron in the $E_{u}$ levels, and a JT effect.  The full band 
dispersion of the square lattice can be written as a superposition of these 
four states (Appendix I of Ref.~\onlinecite{RM8cd}).  It is found that {\it
all states near the VHS's are built up exclusively of $E_{u}$ states}.  On the 
other hand, states near the nodal points have $A_{1g}$ character.  Note that
this explicitly demonstrates that VHS instabilities persist down to `lattices'
as small as a single 2$\times$2 plaquette.

The above calculations are for the weak coupling limit, and are consistent with
the charge ordered states found in Section V.  However, the analysis can be 
repeated in the strong coupling 
limit.  Thus, for an antiferromagnetic arrangement, the (mean field) energies 
are $E(A_{1g})=E(B_{2g})=-t'-\sqrt{\bar U^2+4t^2}$, $E(E_{u})=t'-\bar U$, with 
$\bar U\simeq U/2$.  With parameters\cite{OSP} $t\simeq 325meV$, $U/t\simeq 6$, 
$t'/t\simeq -0.276$, and $J=4t^2/U$, $E(E_{u})-E(A_{1g})\simeq 2t'+J\simeq t/6>
0$, so the first hole would again come from the $E_{u}$ level, although all 
levels are close in energy.  (Note that a smaller $J$ would reverse the order of
the levels but would still lead to an electronic degeneracy.)  Thus, one can 
still have a JT instability in the strong coupling limit, although its nature 
can be significantly modified.  For a somewhat different example, see 
Ref.~\onlinecite{MK1}.

The present JT effect is significantly different from the conventional JT
effect of a CuO$_6$ octahedron, which involves the electronic 
(pseudo-)degeneracy associated with the Cu $d_{z^2}$ and $d_{x^2-y^2}$ orbitals.
Many earlier studies (e.g., Refs. \onlinecite{KAM},\onlinecite{KZ}) have 
proposed that this (pseudo)JT effect plays an important role in cuprate physics.
The present model works even if the $d_{z^2}$ orbitals are completely uncoupled 
(although a residual coupling could enhance the present effects).  The model is 
closer to the large polaron limit\cite{BGood,MKC}, but with a particular choice 
of polaronic coupling dictated by the underlying SO(8) symmetry.

\subsection{The Molecular Jahn-Teller Effect}

As a preliminary to studying the VHJT effect on a lattice, we first look at the
molecular analog, a Cu$_4$O$_4$ molecule with square planar symmetry, $D_{4h}$, 
which corresponds to the $E\otimes (b_1+b_2)$ JT problem\cite{JT1,JT2,EBB}.  
This molecule can display a dJT
effect which bears a striking resemblance to the flux phase.  
Figure~\ref{fig:2} displays the relevant $B_1$ and $B_2$ distortions, assuming
predominantly oxygen vibrations.  The B$_2$ distortion is the oxygen breathing 
mode, while the B$_1$ has the form of the shear wave associated with the LTT 
instability.  

The JT energy can be written $E_{JT}^{(i)}=V_i^2/(2M\omega_i^2)$, 
where $V_i$ is the electron-phonon coupling and $M$ is the ionic mass.  
In terms of $E_{JT}$, there are three classes of solution: 

(i) $E_{JT}^{(1)}\ne E_{JT}^{(2)}$, 

(ii) $E_{JT}^{(1)}=E_{JT}^{(2)}$ but $\omega_1\ne\omega_2$, 

(iii) $E_{JT}^{(1)}=E_{JT}^{(2)}$ and $\omega_1=\omega_2$. 

For case (i) the lowest energy state consists of a static JT distortion
of the mode with larger JT energy only.  
This simple case may apply to the cuprates: it has long been a puzzle why the
oxygen breathing modes do not display the large softening expected\cite{Web} 
near $(\pi ,\pi )$.  On the other hand, the LTT mode is quite soft, and nearly
unstable.  This would follow if $E_{JT}^{(1)}>E_{JT}^{(2)}$.  The results of
Section V are consistent with this possibility.  However, formally the
special cases (ii) and (iii) are more interesting, allowing dynamic solutions 
which are closely related to the flux phase.

Case (iii) reduces exactly to the well-known $E\otimes e$ problem of the
triatomic molecule\cite{HLH,Del,CAM}, and {\it the electronic wave function is 
double valued}: when the molecular motion undergoes a $2\pi$ rotation, the 
electronic wave function picks up an extra factor of $\pi$, the Berry 
phase\cite{CAM,Ber}.  However, in the cuprates the phonon frequencies are very 
different (Section IV), so case (iii) does not apply.  Nevertheless, condition 
(ii) (weak degeneracy) may hold approximately in the cuprates, and it is known 
that inclusion of quadratic vibronic coupling enhances the range of case-(ii) 
degeneracy\cite{Bacc}.  This special case has recently been analyzed\cite{EBB}: 
the dJT effect with $\pi$ Berry phase is preserved, although the
classical motion may be {\it chaotic}.

\begin{figure}
\leavevmode
   \epsfxsize=0.33\textwidth\epsfbox{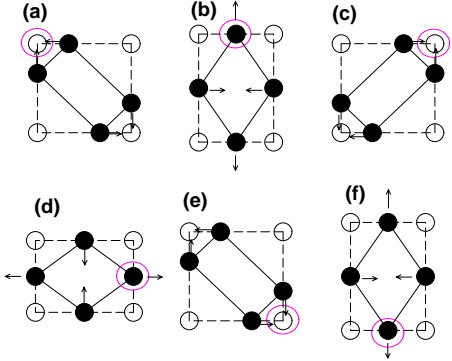}
\vskip0.5cm 
\caption{Dynamic JT rotation mixing $B_1$ and $B_2$ modes of O$_4$.  Large open 
circle represents one hole.  Each time step (a-f) represents one quarter of a
cycle of molecular rotation, so in frame e the molecules have completed one
cycle of rotation with respect to frame a, while the hole has completed only
half a rotation.}
\label{fig:5}
\end{figure}
\par
The origin of the wave function sign change is explained in Fig.~\ref{fig:5}.  
As each oxygen atom rotates about its undistorted position, the hole (large 
circle; a second hole is in the symmetric position) rotates about a different 
origin, the center of the square.  When the individual atoms complete one full 
rotation, the hole has only completed half of a rotation, just as in the 
triangular molecule\cite{HLH}. In analogy with the triatomic molecule, 
the electronic wave function can be kept single valued by 
multiplying it by a phase factor, $e^{i\theta /2}$.  

Thus, the dJT effect with anomalous Berry phase exists for a square
molecule; while some `fine tuning' of the parameters is required, this may
arise from quadratic vibronic coupling\cite{Bacc}.  In the following
subsection, we will see that a similar accidental degeneracy exists for the 
lattice problem, and it is known that the structural transitions in LSCO are
strongly nonlinear\cite{RM8b,LTOtran}.  However, in Section VC it will be seen 
that on a lattice the flux phase can be stabilized over a wider parameter range,
even without nonlinear coupling.

\subsection{Extension to the Lattice}

The $E\otimes (b_1+b_2)$ problem has been extended to the lattice\cite{Mel}
but only the nondegenerate case (i) has been analyzed in detail.  Here, we 
demonstrate that the degenerate case (ii) remains a possibility.  On a lattice, 
the JT coupling becomes
\begin{equation}
H_{JT}=\sum_{i=1,2}H_{JT,i}
\label{eq:41}
\end{equation}
\begin{equation}
H_{JT,i}=-\eta_ie_i\sum_{\vec l}\tau_i(\vec l)-\sum_{\vec q}\sum_mv_i^m(\vec
q)Q_i^m(\vec q)\tau_i(-\vec q),
\label{eq:42}
\end{equation}
where the $m$-sum is over different phonon branches of a given symmetry, and we
include coupling to the elastic strains, with $e_i$ and $\eta_i$ the strain 
component and its coupling to $\tau_i$.  If the phonon Hamiltonian is written
\begin{equation}
H_{ph}={1\over 2}\sum_{i,m,\vec q}M\omega_{i,m}^2Q_i^m(\vec q)Q_i^m(-\vec q),
\label{eq:41b}
\end{equation}
with $M$ an appropriate ionic mass, then the linear coupling to $Q_i$ can be 
replaced by an effective electron-electron coupling by the displaced oscillator 
transformation:
\begin{equation}
\hat Q_i^m(\vec q)=Q_i^m(\vec q)+{v_i^m(-\vec q)\tau_i(\vec q)\over M\omega_{i
,m}^2},
\label{eq:43}
\end{equation}
This transformation approximately decouples the phonons, leaving an interaction
\begin{equation}
H_{int}=-\sum_{i,\vec l}\eta_ie_i\tau_i(\vec l)-{1\over 2}\sum_{i,\vec q}J_i(
\vec q)\tau_i(\vec q)\tau_i(-\vec q),
\label{eq:44}
\end{equation}
with $J_i(\vec q)=\sum_mK_i^m(\vec q)$,
\begin{equation}
K_i^m(\vec q)={v_i^m(\vec q)v_i^m(-\vec q)\over M\omega_{im}^2(-\vec q)}.
\label{eq:45}
\end{equation}
This decoupling is only approximate, since the $\hat Q$'s depend on the 
$\tau$'s, so that they do not obey canonical commutation relations.  It will,
however, be adequate for the present purposes.  At mean field level, 
\begin{equation}
\eta_i<e_i>=\mu_i<\tau_i>,
\label{eq:46}
\end{equation}
with $\mu_i=N\eta_i^2/2c_i^0$ and $c_i^0$ the bare elastic constant.
Letting\cite{Mel} 
\begin{equation}
\lambda_i=\mu_i+\sum_m\bigl(K_i^m(0)-{1\over N}\sum_{\vec q}K_i^m(\vec q)\bigr),
\label{eq:47}
\end{equation}
and assuming only one $\Delta_i$ is non-vanishing, 
the gap equations are
\begin{equation}
\Delta_i=\lambda_i<\tau_i>,
\label{eq:48}
\end{equation}
\begin{equation}
<\tau_i>=-\sum_{\vec k}\Delta_i{f(E_+(\vec k))-f(E_-(\vec k))\over E_+(\vec k)-
E_-(\vec k)},
\label{eq:49}
\end{equation}
with 
$E_{\pm}$ the energy eigenvalues and f(E) the Fermi function.  
If $\lambda_1\ne\lambda_2$, only the mode 
with larger $\lambda_i$ has nonzero expectation value, although both modes can
soften above the transition temperature.  However, when
\begin{equation}
\lambda_1=\lambda_2
\label{eq:50}
\end{equation}
then $\Delta_1=\Delta\cos{\theta_0}$, $\Delta_2=\Delta\sin{\theta_0}$, with 
$\theta_0$ arbitrary.  Equation~\ref{eq:50} is the generalized version of
condition (ii).

\section{Application to the Cuprates}

\subsection{Bond Stretching Modes in LSCO\\ Three-Band Model}

In Section V, we will calculate the mean-field properties of the various CDW's,
combining electron-electron and electron-phonon interaction and including
strong coupling corrections and competition with the Mott gap.  Here we develop
a simple model for estimating the electron-phonon interaction.  We introduce a
three-band model, and apply it to a number of phonon modes
which have been proposed to play a role in the cuprates.  In the following 
subsection, we will reduce the model to a one-band version.

\begin{figure}
\leavevmode
   \epsfxsize=0.33\textwidth\epsfbox{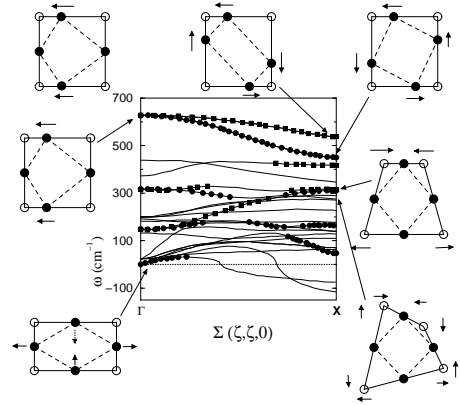}
\vskip0.5cm 
\caption{Phonon modes of the CuO$_2$ planes, including distortion patterns for
several Cu-O bond stretching modes. Clockwise from top left: half-breathing,
breathing, quadrupole, dimer I, dimer II, shear, and ferroelectric modes.  
Filled squares (circles) = $\Sigma_1$ ($\Sigma_3$) branches predominantly 
representing the CuO$_2$ planes.}
\label{fig:1}
\end{figure}

We concentrate predominantly on CuO$_2$ plane phonons associated with in-plane
vibrations, and particularly on Cu-O bond stretching phonons. 
Mott's original picture of a metal-insulator transition involved a loss of
covalency as a lattice is gradually expanded.  Hence, near a Mott transition it
would be expected that bond-length changing phonons might play an important 
role, modulating a crossover from covalent to ionic behavior\cite{Ega}.
Figure~\ref{fig:1} shows the phonon dispersion in La$_2$CuO$_4$ along the $(
\zeta ,\zeta ,0)$ ($\Sigma$) direction, calculated by Wang, et al.\cite{WYK}.  
Shown in bold are the branches which are predominantly associated with in-plane 
Cu-O vibrations.  (Wang, et al. provide the wave functions only along the high 
symmetry axes, $\Gamma$ and $X$; the remainder of the curves are an 
interpolation based on symmetry.  Note the anti-crossing behavior evident on 
several branches.)  The pattern of distortion for a number of modes of interest 
is also shown, including the breathing\cite{Web}, half-breathing\cite{McQ1,Ega} 
(at $(\pi ,0)$), dimer\cite{THMaz}, quadrupolar\cite{KAM}, 
ferroelectric\cite{Ega}, and shear\cite{RMXB} modes.  This by no means exhausts 
all relevant modes.  Among modes which are not modeled are the LTT tilt mode 
and various c-axis modes\cite{TimC}.  As discussed above, we treat the LTT mode 
as a secondary modification of the shear mode instability. 

In order to estimate how strongly these modes couple to the electrons, we 
analyze a simple three-band dispersion for the CuO$_2$ plane, with parameters
$t_0$ ($t_1'$) for Cu-O (O-O) hopping, and $\Delta_0$ ($\Delta_1$) for Cu (O) 
site energy.  The electron-phonon coupling is assumed to arise via modulating
these parameters.  Thus, a change $\delta u$ in the Cu-O bondlength $a$ produces
a change 
\begin{equation}
\delta t=-\beta t_0\delta u/a 
\label{eq:48a}
\end{equation}
in $t_0$ and a change 
\begin{equation}
\delta_0 =\alpha_d\Delta_0\delta u/a
\label{eq:48b}
\end{equation}
in $\Delta_0$ (or $\Delta_1$).  We estimate $\beta\sim 3.5\alpha_t
$\cite{Surv,AAAl} ($\alpha_t=1$) and $\alpha_d$ from a Madelung contribution as 
$\alpha_d=qe^2/\epsilon_0a\Delta_0\sim 0.55$, for $\Delta_0\sim 4eV$, effective 
charge $q=2$ ($O^{2-}$ or $Cu^{2+}$) and dielectric constant $\epsilon_0\sim 5$.
In the following, we will treat the $\alpha$'s as dimensionless parameters of
order unity in order to scale the electron-phonon coupling.  The net change in 
$\Delta$ on any site is then $(n-m)\delta_0$ where $n$ ($m$) is the number of 
near neighbors that move closer to (further away from) the given site.  For a 
frozen lattice distortion of amplitude $\pm\delta u$, we calculate how the net
electronic energy changes.  Model parameters include the position of the Fermi 
level, the value of $t_1'$, and the ratio
$\beta /\alpha_d$.  This last parameter can be thought of as a measure of
covalency: increasing $\Delta$ tends to localize holes on Cu, making the 
material more ionic, while increasing $t_0$ delocalizes holes, increasing
covalency.  We restrict ourselves here to the following parameters: $t_0=1.3eV$,
$\Delta_0=4eV$, $a=1.9\AA$, Fermi level at VHS (in the absence of phonon 
distortion) at a doping $x=0.25$ (to achieve this, 
a large value of $t_1'/t_0=-1.8$ is assumed); and $\alpha_d=\alpha_t=1$.  The 
model neglects possible coupling to the Cu $d_{z^2}$ orbital (important for the 
conventional JT distortions of the quadrupole and breathing modes\cite{KAM}) and
anomalies associated with the near instability of $O^{2-}$ (Section 
V.C).\cite{Bilz}  The two forms of electron-phonon coupling are similar to those
assumed by Yonemitsu, et al.\cite{Yon}; while the numerical estimates are not
identical, we find similar estimates for the coupling to the breathing mode.
Their electron-phonon coupling parameters are, in our notation $\lambda_{\alpha}
=\beta^2t_0/A_{br}\simeq 0.264$, $\lambda_{\beta}=\alpha_d^2\Delta_0^4/t_0A_{br}
\simeq 0.204\alpha_d^2=0.051$, for $\alpha_d=0.5$, and $A_{br}=60.4eV$ from 
Table III.  Yonemitsu, et al.\cite{Yon} estimated $\lambda_{\alpha}=0.28$ and
had no estimate for $\lambda_{\beta}$.

For the shear modes and the modes at $S=(\pi ,\pi )$, the resulting electronic 
hamiltonian can be written in the form $H=$
\begin{equation}
\left(\matrix{\Delta +\delta_1&-t_1e_x&-t_2e_y&0&t_3e_x^*&t_4e_y^*\cr
     -t_1^*e_x^*&\delta_2&-2t'_1c'_-&t_5e_x&0&-2t'_1c'_+\cr
     -t_2^*e_y^*&-2t'_1c'_-&\delta_3&t_6e_y&-2t'_1c'_+&0\cr
     0&t_5^*e_x^*&t_6^*e_y^*&\Delta -\delta_1&-t_7e_x&-t_8e_y\cr
     t_3^*e_x&0&-2t'_1c'_+&-t_7^*e_x^*&-\delta_4&-2t'_1c'_-\cr
     t_4^*e_y&-2t'_1c'_+&0&-t_8^*e_y^*&-2t'_1c'_-&-\delta_5\cr}
\right),
\label{eq:1}
\end{equation}
while for the half breathing mode at $ X=(\pi ,0)$, and the ferroelectric mode, 
$H=$
\begin{equation}
\left(\matrix{\Delta +\delta_1&-t_1e_x&-2it_0s'_y&0&t_3e_x^*&0\cr
     -t_1e_x^*&0&-2it_1's'_ye_x&t_5e_x&0&2it_1's'_ye_x^*\cr
     2it_0s'_y&2it_1's'_ye_x^*&\bar\delta_3&0&2it_1's'_ye_x&0\cr
     0&t_5e_x^*&0&\Delta -\delta_1&-t_7e_x&-2it_0s'_y\cr
     t_3e_x&0&-2it_1's'_ye_x^*&-t_7e_x^*&0&-2it_1's'_ye_x\cr
     0&-2it_1's'_ye_x&0&2it_0s'_y&2it_1's'_ye_x^*&-\bar\delta_3\cr}
\right).
\label{eq:2}
\end{equation}
Here $e_i=exp(ik_ia/2)$, $i=x,y$, $c'_{\pm}=\cos{(k_x\pm k_y)a/2}$ and 
$s'_y=\sin{k_ya/2}$.  For the modes at $S$ (the shear mode), 
$\delta_{i+2}$ = $\delta_i$ ($-\delta_i$) $i=2,3$.  Writing $\delta_i=n_i\delta_
0$ and $t_i=t_0+m_i\delta t_0$, then for oxygen vibrations (the breathing, 
half-breathing, ferroelectric, and quadrupole modes) $m_{i+4}=+m_i$, while for 
modes involving copper motions (the shear and dimers), $m_{i+4}=-m_i$, $i=1
,4$.  For completeness, we include an extra parameter $\bar\delta_3$ in 
Eq.~\ref{eq:2} which is allowed by symmetry, due to second-neighbor relative 
motion; however, to make a uniform comparison with other modes, the value of
$\bar\delta_3$ will be set to zero.  The remaining elements are listed in Table 
I.

\begin{tabular}{||c|c|c|c|c|c|c|c||}        
\multicolumn{8}{c}{{\bf Table I: Phonon Coupling Parameters}}\\
            \hline\hline
Mode & $n_1$&$n_2$&$n_3$&$m_1$&$m_2$&$m_3$&$m_4$ \\
    \hline\hline
1. Breathing&4&0&0&1&1&1&1          \\     \hline
2. Half-Breathing&2&0&2'&1&--&1&--          \\     \hline
3. Quadrupolar &0&0&0&1&-1&1&-1          \\     \hline
4. Dimer I &0&-2&0&1&0&-1&0          \\     \hline
5. Dimer II &0&-2&2&1&-1&-1&1         \\     \hline
6. Dimer III &0&-2&2&1&1&-1&-1         \\     \hline
7. Shear &0&-2&2&1&-1&1&-1          \\     \hline
8. Ferroelectric &0&0&0&1&--&-1&--          \\     \hline
\end{tabular}
\vskip 0.2in

\begin{tabular}{||c|c|c|c||}        
\multicolumn{4}{c}{{\bf Table II: Energy Parameters}}\\
            \hline\hline
Mode & $\gamma$&$X_1$&$X_2$ \\
    \hline\hline
1. Breathing&0&$2X_{10}$&$X_-$          \\     \hline
2. Half-Breathing&0&$X_{10}$&$2(t_0^2-\delta t_0^2)c_{x}$          \\     
        \hline
3. Quadrupolar &0&0&$X_-$          \\     \hline
4. Dimer I &1&0&$X_++4it_0\delta t_0s_{x}$          \\     \hline
5. Dimer II &0&0&$X_++4it_0\delta t_0(s_{x}-s_{y})$         \\     \hline
6. Dimer III &0&0&$X_++4it_0\delta t_0(s_{x}+s_{y})$         \\     \hline
7. Shear &0&0&$2(t_+^2c_{x}+t_-^2c_{y})$          \\     \hline
8. Ferroelectric &0&0&$2(t_0^2-\delta t_0^2)c_{x}$          \\     \hline
9. Flux Dimer I &1&0&$X_-+4it_0\delta t_0c_{x}$          \\     \hline
10. Flux Dimer II &0&0&$X_-+4it_0\delta t_0(c_{x}-c_{y})$         \\     \hline
11. Flux Dimer III &0&0&$X_-+4it_0\delta t_0(c_{x}+c_{y})$         \\     \hline
12. Flux Shear &0&0&$X_--4t_0\delta t_0(s_{x}-s_{y})$          \\     \hline
\end{tabular}
The above calculations refer to static (frozen) phonon modes, and hence do not
give direct information on dJT modes.  In a Hartree-Fock calculation,
dynamic modes can be modelled by imaginary order parameters, and in Section V, 
it is shown that the mean-field decomposition of nearest-neighbor
Coulomb repulsion produces a direct coupling to the OAF (flux phase) mode.  In 
this spirit a similar mean-field phonon flux phase can be constructed 

\vskip 1.3in
\par\noindent
from the above by (a) replacing $\delta t_0$ by $i\delta t_0$ and (b)
making a similar replacement for the $\delta_i$, which involves moving the term
to the off-diagonal ($H_{i,i+3}$) position.

The lattice distortions are found by minimizing the sum of the hamiltonian 
Eq.~\ref{eq:2} and a phonon term, Eq.~\ref{eq:41b}, which may be rewritten as
$H_{ph}=A(\delta u/a)^2$, where $A=\eta M\omega^2a^2/2$, and $A$ and $\omega$
are listed in Table III.  Strictly speaking, the phonon frequencies should be
bare values, neglecting electron-phonon coupling, but for present purposes we
use calculated\cite{WYK} or experimental\cite{PintR} values for the undoped
insulator La$_2$CuO$_4$.  [For the shear mode, $A=(C_{11}-C_{12})a^2c$, where
$(C_{11}-C_{12})/2=99GPa$ [\onlinecite{SSNN}] and $c$ is the c-axis lattice 
constant.]  The parameter $\eta$ is one if the distortion is along the $X$ or
$Y$ axis only, two if it is along both, and $M$ is the mass of the moving ion,
oxygen or copper (or the appropriate average for the ferroelectric mode).
These equations must in general be solved numerically, but some insight can be 
gotten for
the special case $n_2=n_3=t'_1=0$, for which the four non-zero eigenvalues are 
given by 
\begin{equation}
E_{i\pm}={\Delta_0\pm\sqrt{\Delta_0^2+4W_i^2}\over 2}, 
\label{eq:33a}
\end{equation}
with $i=\pm 1$,
$W_{\pm}^2=W_1^2\pm \sqrt{X_1^2+|X_2|^2}$, $W_1^2=4(t_0^2+(1-\gamma /2)\delta 
t_0^2)$ (except for the half breathing and FE modes, for which $W_1^2=2(t_0^2[1+
2s'^2_y]+\delta t_0^2)$) and the $X$'s and $\gamma$'s are listed in Table II, 
with $s_i=sin(k_ia)$ (and recall $c_i=cos(k_ia)$) $X_{\pm}=2(t_0^2\pm
\delta t_0^2)(c_{x}+c_{y})\mp 2\gamma\delta t_0^2c_{y}$ 
and $X_{10}=2E\delta_1+
4t_0\delta t_0$ (note that the solution is only implicit if $\delta_1\ne 0$).
The generic form of $X_2$ is $X_2=t_1t_5e_x^2+t_3^*t_7^*e_x^{*2}+t_2t_6e_y^2+
t_4^*t_8^*e_y^{*2}$.  Note that $E_{i+}$ is the antibonding band which is 
approximately half filled.

\begin{figure}
\leavevmode
   \epsfxsize=0.33\textwidth\epsfbox{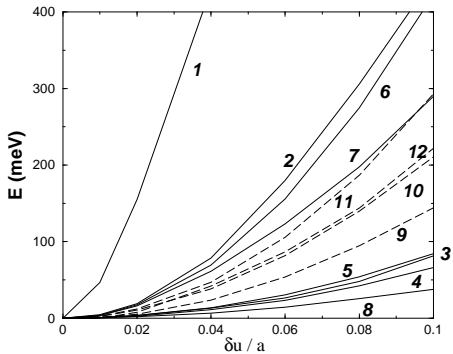}
\vskip0.5cm 
\caption{Energy of various phonon modes of CuO$_2$ planes, assuming $\alpha_d=1$
and $\beta$ = 3.5, for static (solid lines) or flux phases (dashed lines).  The
phonon modes are numbered according to Table II.}
\label{fig:2a}
\end{figure}

\begin{figure}
\leavevmode
   \epsfxsize=0.33\textwidth\epsfbox{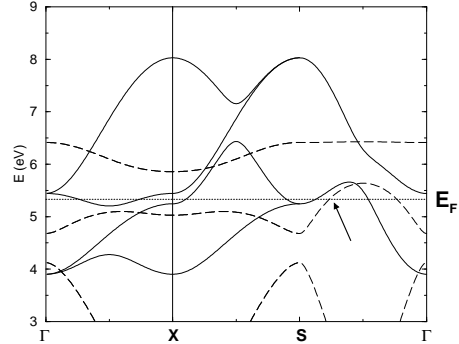}
\vskip0.5cm 
\caption{Three-band dispersion of the CuO$_2$ planes, in the presence of a
frozen breathing mode (dashed lines) or half-breathing $(\pi /a,0)$ mode (solid 
lines) modulation, assuming $\beta /\alpha_d=3.5$, and $\delta u/a$ = 0.02 
(breathing) or 0.04 (half breathing).  The arrow indicates the breathing mode
hole pocket near $(\pi /2a,\pi /2a)$.  Brillouin zone special points are $\Gamma
=(0,0)$, $X=(\pi /a,0)$, $S=(\pi /a,\pi /a)$.}
\label{fig:4}
\end{figure}

Figure~\ref{fig:2a} shows the calculated electronic energy shifts up to large 
distortions $\delta u/a=0.1$ (local distortions of up to about 5\% are reported 
in EXAFS measurements\cite{Bian}).  A number of interesting features can be 
noted.  (1) All the distortions lead to highly nonlinear shifts of the 
electronic energy.  (2) While the 
breathing mode has the strongest coupling (as originally predicted by 
Weber\cite{Web}), the half-breathing mode also has a large coupling, since 
$\delta_1$ produces a large splitting at the VHS, Fig.~\ref{fig:4}.
(3) The energy shifts correspond to significant phonon softening, but when the 
quadratic phonon energy is included, it is found that only the breathing and
(marginally) half breathing mode become unstable -- the additional softening due
to electron-electron coupling is discussed in Section V.  Instabilities of the
dimer modes were found earlier\cite{THMaz}, but only in strong nesting 
conditions -- half filling with $t_1'=0$. (4) A problem for a pure phonon-flux
phase should be noted: the symmetric dimer III mode has a larger energy lowering
than the asymmetric ($c_x-c_y$) dimer II mode.

\begin{figure}
\leavevmode
   \epsfxsize=0.33\textwidth\epsfbox{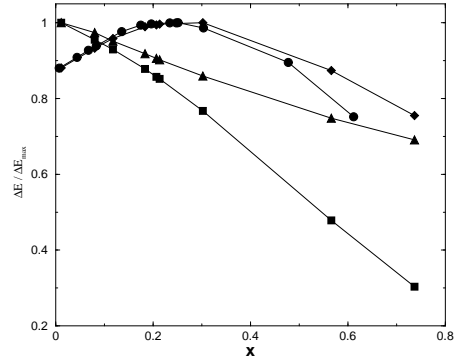}
\vskip0.5cm 
\caption{Doping dependance of the frozen phonon energy lowerings, associated
with the breathing mode (circles), half-breathing $(\pi /a,0)$ mode (squares), 
shear mode (diamonds), and dimer III mode (triangles), all normalized
to their maximum amplitude.  The half-breathing and shear modes have strongest
coupling near the VHS. $\delta u/a$ = 0.04 for the breathing mode, 0.08 for the
others.}
\label{fig:41}
\end{figure}

The doping dependence of the energy lowering is shown in Figure~\ref{fig:41}.
The results are in good agreement with experiments\cite{McQ1} on both LSCO and 
YBCO and with LDA calculations\cite{Falt} which find that the half-breathing 
mode softens more than the $(\pi ,\pi )$ breathing mode.  This result is 
surprising since the breathing mode has stronger coupling. However, this mode 
has an unusual frustration effect: it opens a complete gap at the Fermi level, 
Fig.~\ref{fig:4}, so the Fermi level can only fall in the gap if the band is 
filled, $x=0$.  To accomodate additional holes, the Fermi level must shift {\it 
below} the gap, reducing the electronic energy lowering.  Hence optimal doping 
for this mode is at half filling, and {\it while there is a large electronic 
softening of the breathing mode at the VHS, there is an even larger softening 
near half filling, so a doping dependence would show a hardening of the mode.}  
Indeed, such a hardening is observed in the nickelates\cite{TraNO}.  This result
persists in the presence of strong correlation effects, Section V.
For the half-breathing mode correlation effects should enhance the already 
large softening of the mode with doping, in good agreement with experiment.

\begin{figure}
\leavevmode
   \epsfxsize=0.33\textwidth\epsfbox{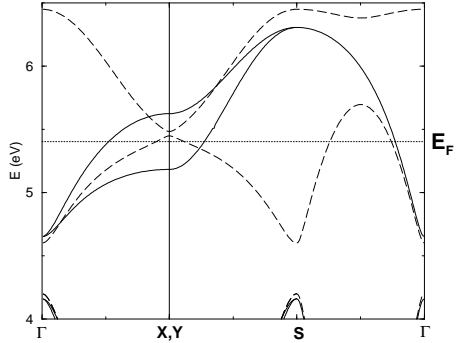}
\vskip0.5cm 
\caption{Three-band dispersion of the CuO$_2$ planes, in the presence of a
frozen dimer II mode (long-dashed lines) or shear mode (solid lines) modulation,
assuming $\beta /\alpha_d=3.5$, and $\delta u/a$ = 0.04.}
\label{fig:5a}
\end{figure}

\begin{figure}
\leavevmode
   \epsfxsize=0.33\textwidth\epsfbox{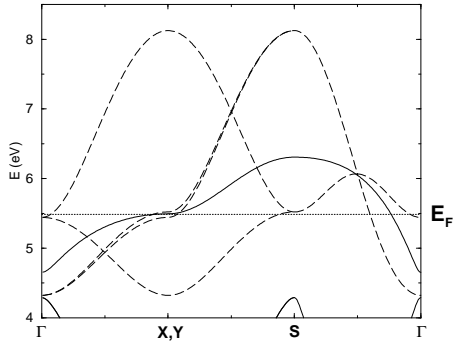}
\vskip0.5cm 
\caption{Three-band dispersion of the CuO$_2$ planes, in the presence of a
frozen ferroelectric (solid lines) or quadrupole mode (dashed lines) modulation,
assuming $\beta /\alpha_d=3.5$, and $\delta u/a$ = 0.08.}
\label{fig:6}
\end{figure}

Figures~\ref{fig:4}-\ref{fig:6} compare the dispersions of several of the 
modes, near the upper (Cu-like) band of the three-band model.  For all cases,
the magnitude of electronic energy lowering can be directly correlated with the 
size of the gap at the VHS ($X$-point).  Note that the gap is largest for the 
breathing mode, but the Fermi level lies below the gap if $x\ne 0$.  For the
modes at $\Gamma$, there is no unit cell doubling, but the dispersion is
different along $X$ and $Y$, leading to an effective splitting of the VHS's, 
Figs.~\ref{fig:5a}-\ref{fig:6}.

For the half-breathing mode we find that the softening is via the on-site Cu 
energy, $\Delta$.  Making $\Delta$ inequivalent on alternative rows of Cu's 
along the $X$-axis (parallel to the oxygen distortion) leads to a large 
splitting of the VHS's.  Figure~\ref{fig:4} shows that the gap is smaller than 
for the full breathing mode, but is clearly centered at the VHS.  

\subsection{Bond Stretching Modes in LSCO\\ One-Band Model}

To include electron-electron coupling, it is convenient to first reduce the 
above results to a one-band model.  The parameters of the one-band model can 
readily be related to those of the three band model by expanding the approximate
eigenvalues of Eq.~\ref{eq:33a} as $E_{i+}=\Delta_0+W_i^2/\Delta_0$.  In this
way, it is found that the effective Cu-Cu hopping parameter is $t=t_0^2/\Delta_
0$ [with the parameters assumed above, this would give $t=0.42eV$, somewhat
larger than the value 0.326eV estimated from photoemission], with phonon 
distortion parameters $\delta t/t=2\delta t_0/t_0$ and $\bar
\delta_1=\delta_0+\delta t$.  Linearizing the results in $\delta t_0$, it is 
readily seen that the flux dimer II mode has the symmetry of the 
conventional flux phase, and it is found that the quadrupole and ferroelectric 
modes do not couple at the one band level, while the breathing mode couples 
only via $\bar\delta_1$.  The electronic dispersion can be solved:
\begin{equation}
E=-4t'c_xc_y\pm\sqrt{(m_1\bar\delta_1)^2+|X_1+X_2|^2},
\label{eq:4a}
\end{equation}
with $m_1$ and $X_2$ listed in Table III, and $X_1=-2t(c_x+c_y)$.  The
result for the half breathing mode is distinct:
\begin{equation}
E=-2tc_y\pm\sqrt{(2\bar\delta_1-2\bar{\delta t}c_y)^2+4(t+2t'c_y)^2c_x^2}.
\label{eq:5}
\end{equation}
The resulting dispersions are in excellent agreement with the
corresponding antibonding band dispersions of the three band model.  

For the half breathing mode the term in $\bar{\delta t}$ follows from the 
$\bar\delta_3$ term in Eq.~\ref{eq:2}: for a phonon distortion along $X$, the 
hopping along $Y$ is modified along alternate rows, at $t\sim t_0^2/(\Delta_0\pm
\bar\delta_3)$.  Such a term has been previously considered by Shen, et 
al.\cite{ZXph}.  In the present comparisons of different modes, the role of such
a term has been neglected (by setting $\bar\delta_3=0$).

\begin{tabular}{||c|c|c|c|c||}        
\multicolumn{5}{c}{{\bf Table III: One Band Model Parameters}}\\
            \hline\hline
Mode & $m_1$&$X_2$&$\omega$(meV)&A(eV) \\
    \hline\hline
1. Breathing&4&0&66&60.4          \\     \hline
2. Half-Breathing&2&--&83&47.8          \\     
        \hline
3. Quadrupolar &0&0&56&43.5          \\     \hline
4. Dimer I &0&$-2i\delta ts_{x}$&38&39.7          \\     \hline
5. Dimer II &0&$-2i\delta t(s_{x}+s_{y})$&38&79.5         \\     \hline
6. Dimer III &0&$-2i\delta t(s_{x}-s_{y})$&38&79.5         \\     \hline
7. Shear &0&$-2\delta t(c_x-c_y)$&--&59          \\     \hline
8. Ferroelectric &0&0&77&32.8          \\     \hline
9. Flux Dimer I &0&$-2i\delta tc_{x}$&38&39.7          \\     \hline
10. Flux Dimer II &0&$-2i\delta t(c_{x}-c_{y})$&38&79.5         \\     \hline
11. Flux Dimer III &0&$-2i\delta t(c_{x}+c_{y})$&38&79.5         \\     \hline
12. Flux Shear &0&$+2i\delta t(s_{x}-s_{y})$&--&59          \\     \hline
\end{tabular}
\vskip 0.2in

\section{Results in Strong Coupling Limit}

\subsection{Electron-Electron Coupling}

In one-dimensional metals, the CDW instability involves both electron-electron
and electron-phonon coupling.  Here we show that the same is true in the present
two-dimensional problem: electron-electron interaction actually dominates the 
electron-phonon coupling in the cuprates.  A nearest neighbor Coulomb 
interaction $V$ contributes to all three components of pseudospin, 
Eqns.~\ref{eq:12b}-\ref{eq:12a}\cite{MK1}; it is estimated\cite{BS} that $V
\simeq 0.2-0.3eV$. 

\begin{figure}
\leavevmode
   \epsfxsize=0.33\textwidth\epsfbox{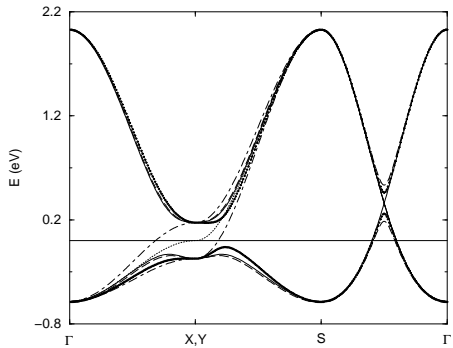}
\vskip0.5cm 
\caption{Energy dispersion in the presence of pseudospin ordering, 
Eq.~\protect\ref{eq:8}, for $t$ = 0.326eV, $t^{\prime}=-0.276t$, and $R_{x0}=
R_{y0}=R_{z0}$ = 0.1eV (short dashed line); or $R_{i0}=0.1732$eV and the other 
two $R$'s = 0, for $i$ = $x$ (dot-dashed line), $y$ (solid line), or $z$ (long 
dashed line); or all $R_i$'s = 0 (dotted line).}
\label{fig:3}
\end{figure}

The electronic part of the Hamiltonian can be written $H=H_{el}+H_{1}$
where the electronic kinetic energy is 
\begin{equation}
H_{el}=\left(\matrix{E_1&E_0\cr
     E_0&E_1\cr}\right),
\label{eq:6}
\end{equation}
with $E_0=-2t(c_x+c_y)$, $E_1=-4t^{\prime}c_xc_y$, and $t$ ($t^{\prime}$) is the
(next) nearest neighbor hopping parameter.  At mean field, the effects of both 
$V$ and the on-site Coulomb interaction $U$ can be included, leading to the 
Hamiltonian matrix
\begin{equation}
\left(\matrix{R_{z\sigma}+E_1&E_0+R_x+iR_y\cr
     E_0+R_x-iR_y&-R_{z,\sigma}+E_1\cr}\right),
\label{eq:7}
\end{equation}
with eigenvalues 
\begin{equation}
E_{\pm,\sigma}=E_1\pm\sqrt{R^2+E_0^2+2R_xE_0}, 
\label{eq:8}
\end{equation}
where $R^2=R_x^2+R_y^2+R_{z,\sigma}^2$, Fig.~\ref{fig:3}.  In 
Eqs.~\ref{eq:7},\ref{eq:8}, we take $R_x=R_{x0}\tilde\gamma$, $R_y=R_{y0}
\tilde\gamma$, and $R_{z,\sigma}=R_{z0}+\sigma Um_z$, where the $R_{i0}$'s 
are constants, $\sigma =\pm 1$ is the spin, and $m_z$ is the average 
magnetization per site in the antiferromagnetic phase induced by $U$.  Parameter
values are given in Section IIIA.  The 
conventional\cite{Affl} flux phase dispersion is recovered from Eq.\ref{eq:8} 
for the special choice of parameters: $t'=R_{x0}=R_{z,\sigma}=0$ and $R_{y0}=2
t$ (compare Eq. \ref{eq:11}).

Note that while the interaction $H_1$ resembles $H_{vib}$ of Eq. \ref{eq:5}, 
there are a number of differences.  First, while the term involving $T_y$ is
generated dynamically in the molecular problem, here a term in $R_y$ appears
explicitly.  
On a single plaquette, the three solutions are degenerate (the eigenvalues 
depend only on $R$, and not on the individual $R_i$'s), but intercell hopping 
and $k$-dependent coupling lead to slight dependence of the dispersion on the
individual $R_{i0}$'s.  (Note that $R_x$ simply renormalizes $t$ to have
unequal values in the $x$ and $y$ directions. The effects of this term have been
recently analyzed\cite{HaM,VV}.) The solutions remain degenerate at the VHS, for
equal $R_{i0}$'s.  

Introducing the polarization
\begin{equation}
\Pi_{\vec k,\sigma}=-{f(E_{+,\sigma})-f(E_{-,\sigma})\over E_{+,\sigma}-
E_{-,\sigma}},
\label{eq:8a}
\end{equation}
the gap equations can be written
\begin{equation}
\sum_{\sigma}\int_{\vec k}\Pi_{\vec k,\sigma}\Lambda_i=1,
\label{eq:8b}
\end{equation}
for $i$ including all the non-vanishing gap contributions.  The most general
case has all four gap parameters, $R_{x0},R_{y0},R_{z0},m_z$, nonvanishing, 
with corresponding $\Lambda_1=\lambda_1(\tilde\gamma +E_0/R_{x0})\tilde\gamma$, 
$\Lambda_2=\lambda_2\tilde\gamma^2$, $\Lambda_3=\lambda_3(R_{z,
\sigma}/R_{z0})$, and $\Lambda_4=\lambda_4(1+\sigma R_{z0}/Um_z)$.  For the 
extended Hubbard model, the $\lambda$'s are\cite{MK1}: $\lambda_1=\lambda_2=8V$,
$\lambda_3=16V-2U$, and $\lambda_4=2U$.  (In the areas of overlap, these gap 
equations agree with those found by Nayak\cite{Nay}.  However, a strong coupling
calculation\cite{PhPh} has suggested that only an attractive near-neighbor
coupling would stabilize the flux phase.)

In the absence of on-site $U$, the near-neighbor interaction $V$ favors the CDW,
$R_z$.  The energy lowering is exactly twice that of the other components, the 
factor of two arising from spin (since the $x$ and $y$ terms involve hopping, 
only terms with the same spin on both neighbors contribute; the $z$ term has no 
such limitation).  This is counteracted by strong coupling effects, since
the CDW involves an imbalance of Cu site occupancies.  Approximating $V=t$,
$U=6t$, $8V>16V-2U$, favoring the flux and shear phases.  Because the
couplings are not symmetrical (Eq.~\ref{eq:8}), these modes have slightly 
different gaps, $R_{x0}=340meV$, $R_{y0}=320meV$, $R_{z0}=8.4meV$.  However, 
$2U>8V$, so in a purely electronic model the antiferromagnetic phase dominates,
both at half filling and in the doped case.  In the following
subsections, it will be shown that electron-phonon coupling can reverse this 
situation in the doped cuprates, and also brings the CDW energy closer to that
of the other phases.

\subsection{Electron-Phonon Coupling}

Combining the above results with electron-phonon coupling, we limit discussion
to the three modes of primary interest.  
In the one-band model, the linear electron-phonon coupling has exactly the
same form as the $V$ electron-electron term in Eq.~\ref{eq:7}.  Thus, the free 
energy $F$ is a function only of the {\it sums} of the electron-electron and
electron-phonon gaps, leading to a simplification of the gap equations.  For
example, for the CDW $F=F(R_{z0}+4\delta_1)$, so 
\begin{equation}
<{\partial F\over\partial\delta_1}>=4<{\partial F\over\partial R_{z0}}>.
\label{eq:5b}
\end{equation}
Since the averages are just those evaluated in the gap equations, they can be
replaced by the derivatives of the quadratic terms, yielding:
\begin{equation}
{\delta_1\over R_{z0}}={8\alpha_d^2\Delta_0^2\over A(16V-2U)}\simeq 0.22
\label{eq:5c}
\end{equation}
for $\alpha_d=0.5$ and the other parameters as given above.  For the shear mode,
the equivalent result is  
\begin{equation}
{\delta t\over R_{x0}}={\beta^2t^2\over AV}\simeq 0.068,
\label{eq:5d}
\end{equation}
with $\alpha_t=1$; an identical result holds for the flux phase, with 
appropriate $A$, yielding $\delta t/R_{y0}\simeq 0.05$.  These are the analogs 
to Eq. \ref{eq:43}.  They
are weak coupling results and should be modified when the gaps become comparable
to the phonon frequencies.  Thus, (1) there is substantial phonon coupling,
but the transitions appear to be predominantly electronically driven; (2) the
largest correction is for the CDW mode, although the correction is not large
enough to make the CDW as unstable as the other two modes (if $\alpha_d$ were
twice as large as estimated, all three modes would be approximately degenerate).
Finally, (3) near half filling the corrections are not large enough to make any 
of the structural anomalies competitive with the AFM instability. 
This result is consistent with the finding of Hsu, et al.\cite{Hsu}, that the
flux phase is unstable against magnetic order near half filling.
It remains possible that additional couplings could tip the balance toward 
structural distortions, particularly in the doped materials.  Certainly, there 
are strong indications of deviations from Migdal theory\cite{Piet}, and earlier
calculations of the LTO and LTT instabilities found strong 
nonlinearities\cite{LTOtran,RM8b}.  In the following subsection we explore 
another possibility, a particular form of correlated hopping.

\subsection{Unconventional Coupling Associated with O$^{2-}$}

The O$^{2-}$ ion is known to be inherently unstable, being stabilized in a
solid by the Madelung potential of surrounding ions.  This near instability
has been suggested to be a driving force in ferroelectric transitions\cite{Bu1} 
in perovskites and in cuprate superconductivity\cite{Bu2}.  One of the principal
manifestations of this near instability is the large change in ionic radius on 
doping O$^{2-}$ to O$^-$.\cite{RMXB}  While this anomaly should affect the
electronic properties in a number of ways, we will here explore only one aspect,
a correlated hopping.

When a hole is {\it localized} on a single oxygen, the shrinkage of its radius
allows the adjacent coppers to approach much more closely.  This would enhance 
the corresponding hopping probability, $t$, except that the presence of the hole
inhibits other holes from hopping to the same site.  If, however, the first hole
hops away, it will take the local lattice some time to relax back, and in that
interval there will be an enhanced probability of another hole hopping onto
the oxygen.  In a one band model, this would correspond to correlated hopping
between adjacent Zhang-Rice singlets, or effectively, between adjacent coppers.
This hopping adds a term 
\begin{equation}
H_{c.h.}=-t_{c1}\sum_{<i,j>,<i',j>,i\ne i'}c^{\dagger}_{i\sigma}c_{j\sigma}
c^{\dagger}_{j\sigma '}c_{i'\sigma '}
\label{eq:8c}
\end{equation}
to the Hamiltonian.  Such a term has been condidered by Nayak\cite{Nay}; 
for related models, see Hirsch and Marsiglio\cite{HiM}.  This term
contributes $96t_{c1}$ to all three coupling parameters $\lambda_i$, $i=1,3$.
From the present considerations, a value $t_{c1}=-2x\beta_{loc}\beta t\delta 
u_0/a\sim xt$ can be estimated, where $\delta u_0\sim -0.3\AA$ is the shrinkage 
in radius\cite{RMXB} on going from $O^{2-}$ to $O^-$.  The factor $x$ arises
because holes only move preferentially onto the oxygens for doping beyond half
filling, $x>0$, while $\beta_{loc}\le 1$ is a parameter introduced to describe
the degree of localization of the hole on a single oxygen -- predominantly due
to polaronic effects.  This factor is related to the delicate issue of the 
crossover from ionic to covalent behavior: $\beta_{loc}=1$ corresponds to the 
ionic limit.  The $t_{c1}$ correction is so large that even a significantly
covalent correction $\beta_{loc}<1$ would allow the structural 
distortions to overcome the AFM state -- but only for $x>0$.  

\subsection{Numerical Results}

\begin{figure}
\leavevmode
   \epsfxsize=0.33\textwidth\epsfbox{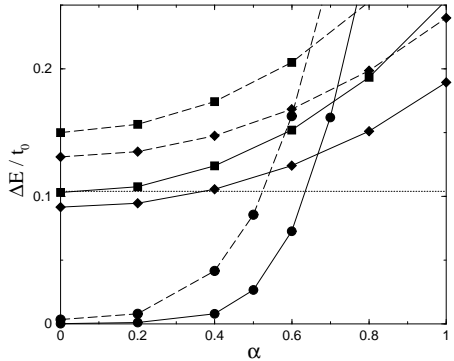}
\vskip0.5cm 
\caption{Binding energy of various modes -- site CDW (circles), bond CDW 
(squares) and flux phase (diamonds), for $\beta_{loc}$ = 0 (solid lines) or 0.1
(dashed lines); horizontal dotted line = energy of doped antiferromagnetic 
phase.  For all curves $x=0.25$.}
\label{fig:1xa}
\end{figure}

Here we describe our numerical results.  The gap equations follow from 
minimizing the mean-field free energy
\begin{equation}
F=\sum_{k,i=\pm}E_if(E_i)-TS+Nf_0,
\label{eq:7x1}
\end{equation}
where $S$ is the entropy, $N$ the total number of electrons, and
\begin{equation}
f_0={R_{x0}^2+R_{y0}^2\over 4V}+{R_{z0}^2\over 8V-U}+\sum_{i=1,3}A_i({\delta u_i
\over a})^2+U(m_q^2+{x^2\over 4}).
\label{eq:7x2}
\end{equation}
The $E_i$ are solutions of Eq.~\ref{eq:8} with phonon coupling included by the
substitutions: $R_{z0}\rightarrow R_{z0}+4\bar\delta_1$, $R_{i0}\rightarrow 
R_{i0}+2\delta t$, $i=x,y$.
In Figure~\ref{fig:1xa} we gradually turn on the phonon coupling, by varying the
$\alpha$ factors from zero (no phonon coupling) to 1, covering the expected 
range of coupling.  Here we study the coupling near the VHS, $x=0.25$.  What is 
plotted is the self-consistent condensation energy.  For weak phonon coupling, 
the CDW has the weakest coupling, due to the on-site Coulomb repulsion.  
However, it has the strongest intrinsic coupling to phonons, so as $\alpha_d$ 
is increased, it grows fastest, and can actually cross the other
modes.  The dotted line is the energy of the simple doped antiferromagnet (also
at $x=0.25$).  It can be seen that for reasonable phonon coupling, the ground
state is paramagnetic.  The dashed lines show that even moderate correlated
hopping ($\beta_{loc}=0.1$) significantly enhances the binding energies, and
shifts the degeneracy points of the CDW with the other modes to lower values of
$\alpha$.

\begin{figure}
\leavevmode
   \epsfxsize=0.33\textwidth\epsfbox{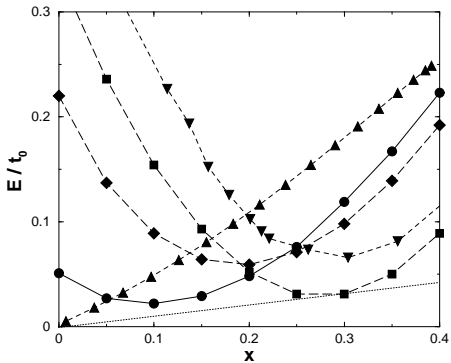}
\vskip0.5cm 
\caption{Doping dependence of binding energies of phonon modes: site CDW 
(circles), bond CDW (squares), and flux phase (diamonds), assuming 
all mode $\alpha$'s = 0.6, and $\beta_{loc}=0.1(x/0.25)$.  Also shown is the 
doped antiferromagnet (triangles) and ferromagnet (inverted triangles)
[\protect\cite{MKII}].  The dotted line is the tangent 
construction for phase separation.  For convenience in viewing, an energy shift 
$\delta E=2.1(1-x)t_0$ has been added to all curves.}
\label{fig:1xb}
\end{figure}

In Fig.~\ref{fig:1xb} the doping dependence of the binding energies is 
illustrated for a representative $\alpha_i=0.6$ ($i=d,t$), compared to that of 
the doped antiferromagnet and ferromagnet\cite{MKII}.  All three CDW modes have 
a parabolic binding energy vs doping: for the bond-CDW and flux modes, the 
energy minima are near the VHS doping, while the site-CDW minimum is shifted to 
lower doping by the frustration effect discussed above (the minimum is no longer
at $x=0$, due to the strong correlation effect).  A by now 
familiar\cite{RM3,MKK} Van Hove induced phase separation could arise between the
antiferromagnetic insulator at half filling and {\it any} of the CDW modes.  For
the present parameter values, the lowest energy state involves the bond CDW at 
$x\sim 0.27$.  Note in particular that the bond CDW is more stable than the
ferromagnetic phase\cite{MKII}.  This is consistent with the numerical
results of Yonemitsu, et al.\cite{Yon}, who found a crossover from magnetic to
dielectric polarons in the low doping regime.  Remarkably, all three CDW 
curves approximately converge (weak JT degeneracy) at a doping, $x\sim 0.21$, 
close to the VHS.  However, this crossover point is not directly observable, 
due to the phase separation.   At this time we have not made a detailed study of
the parameter dependence of these results, but the bond-CDW appears to be most
stable.  In particular, (1) $\alpha_d$ should probably be less than $\alpha_t$;
and (2) our estimate for $V$ may be a little large, but $t_{c1}$ is probably too
small, so $V$ can be reduced and $t_{c1}$ increased to keep a fixed coupling to 
the shear mode.  Both of these corrections would act to further weaken the 
site-CDW.  The results are quite intriguing.  For a reasonable
estimate of the parameters there is phase separation between the 
antiferromagnetic insulator and a paramagnetic state associated with LTT-like
distortions, while the CDW and flux phase states are close in energy.

These results are in good agreement with the numerical calculations of 
Yonemitsu, et al.\cite{Yon}.  In their work $V$ was not included, so their
phonon anomalies are of the breathing mode type, but they see clear evidence for
nanoscale phase separation between antiferromagnetic background and dielectric
polarons.

The present approach of turning on the electron-phonon interaction last
(Fig.~\ref{fig:1xa}) makes the resulting near-degeneracy seem rather accidental.
It is more natural to reexpress the result: for {\it both}
electron-electron and electron-phonon coupling, the flux and bond-CDW phases
are nearly degenerate, while the site-CDW is more strongly coupled.  However,
the on-site Coulomb repulsion opposes site-CDW formation, making the three modes
nearly degenerate in energy.

\section{Discussion: Future Directions}

The present paper establishes the framework for analyzing CDW's in the cuprates,
and Berryonic matter in a number of related materials.  Clearly much work 
remains to be done.  This includes:

(1) From the associated susceptibilities the modifications to the electronic
dispersion can be calculated and compared with the kink\cite{PEkink,PEk2} seen
in photoemission.  While phononic contributions appear to be 
important\cite{ZXph}, Valla\cite{Val} has noted that the linear frequency 
dependence of the imaginary self energy extends to much too high frequencies to
be solely due to strong electron-phonon coupling.  This is consistent with the
present results, that the CDW has a strong electron-electron component.  Hence,
both components will be important in interpreting the photoemission dispersion,
with proximity to a VHS providing a marginal Fermi liquid-like\cite{RM4} 
background and phonon coupling the kink in the dispersion.  

(2) The manner in which the $y$-component couples to the Hamiltonian is markedly
different in the molecular and lattice versions of the theory: in the molecular
version, the direct electron-phonon coupling is to the $x$ and $z$ components
($B_1$ and $B_2$ modes), with the dJT term generated dynamically -- from
the phonon kinetic energy term.  In the lattice, the flux phase term arises
directly from electron-electron coupling via $V$, and corresponding terms are
allowed in the Hartree-Fock expansion of the electron-phonon coupling.  It
remains to be seen what role phonon kinetic energy plays in the lattice
problem.  [One effect will presumably be the kink in the electronic dispersion
when $E=\hbar\omega_{ph}$.]

(3) Fluctuations of the CDW's (or of the stripes) should appear as new low 
frequency (or pinned) phason modes\cite{EBB}.  It is possible that such modes
have been observed in microwave measurements.\cite{phaso}

(4) It remains to work out the detailed doping dependence, in terms of stripes.

(5) It will be important to provide detailed calculations showing how the shear
mode couples to the various bond bending modes in the different cuprates.  A 
related issue: are there tilt distortions in the flux phase?  (if so, then
experimental evidence for local LTT order could really be associated with the
flux phase.)

(6) One puzzling feature is the role of the LTO phase.  Near optimal doping it
would seem to be predominantly associated with local LTT order, but it seems to
be a uniform phase near half filling.  In principle, its presence could be 
accidental, particularly since it is only present in LSCO, but it does provide
an easy axis for orienting the spins, and hence may be involved in the stripe
crossover from vertical to diagonal.  (Also, the pseudogap seems to follow $T_{p
g}(x)=2T_{LTO}$.\cite{Surv})

(7) A very similar model should apply to the nickelates, and extensions can be
made to other forms of Berryonic matter.

(8) There remains the problem of the competition of density wave order with
superconductivity, and the possible roles of magnetic fluctuations.

\section{Conclusions}

This results of this paper have bearings on four separate issues.  First, on a
purely formal level, the paper 
presents an alternative approach (phonon flux phases) to the dJT effect 
in solids, revealing previously unsuspected analogies between flux phases and 
Berryonic matter.  
Secondly, The approach is applied to the cuprates, which are an {\it a 
priori} unlikely candidate for JT effects.  Nevertheless, a `hidden' JT 
degeneracy is found, and a numerical estimate suggests that the cuprates are
close to the weak degeneracy point of a square molecule, which should enhance 
the possibility of a dynamic flux phase.  These calculations can be considered
a generalization of the results of Yonemitsu, et al.\cite{Yon}, by (a) reducing
the problem to a one band model, (b) including a nearest neighbor Coulomb
repulsion which enhances the scope of CDW-like instabilities, and (c) providing
a mean-field underpinning for the numerical calculations of local electronic
phase separation.  
Third, the paper explores the relative strength of different CDW-like 
distortions in the cuprates, and suggests an explanation for strong coupling to 
the LTT and half-breathing modes.  The role of these
distortions in stabilizing charged stripes is discussed in an Appendix. Finally,
the results offer strong support for the general picture of stripe phases as
stabilized by VHS-induced ordering.  A similar
approach should be applicable to other systems, and in particular, to nickelates
where the coupling to the CDW/breathing mode is known to be stronger.

For the cuprates, we find that the structural distortions are close to the 
dJT degeneracy, but that the shear and flux
phases are more unstable than the CDW.  Hence, one would expect significant
softening of the breathing modes, but instability in either the shear or flux 
phases.  These expectations are borne out in experiments on the cuprates, 
with the shear mode coupling to a (local) LTT order.
The near degeneracy of the shear and flux phases is consistent with experimental
evidence for both modes.
Perhaps the most interesting phase is the flux phase, which
is closely related to a chaotic dJT phase.  

Recently, dJT phases have been proposed in a number of exotic materials,
including cuprates\cite{RM8cd}, buckyballs\cite{Auer} and 
manganates\cite{AJM}.  To describe the dJT phase of buckyballs, the 
concept of Berryonic matter was introduced\cite{Berr}.  In this model, no 
attempt was made to accurately model Buckyball solids.  Instead, a lattice of 
dJT molecules was assumed, basing their properties on the known anomalies
associated with triatomic molecules.  Since the square molecule, 
Fig.~\ref{fig:2}, has the same dJT anomaly as the triangular model 
when $E_{JT}^1=E_{JT}^2$, and this degeneracy persists on a square lattice
(Section IIIC), the present model should be an excellent starting point for
studies of Berryonic matter.  Moreover, the proximity of the cuprates to the
Berryonic limit should stimulate interest in this unusual state of matter.

\appendix 
\section{Phonons on Stripes}

In the Van Hove model, charged stripes (or domains) are 
stabilized at a particular density by a VHS driven ordering instability, most
likely a CDW\cite{RM3} as analyzed here.  Originally, this distinguished the
VHS stripe model from magnetic models, in which the charged stripes were
structureless domain walls between antiferromagnetic domains.  However, in 
1995, it was found\cite{Tran} that the hole density on the charged stripes
in LSCO:Nd is considerably less than one per copper, and the magnetic models 
were modified to include CDW order on the charged stripes\cite{CDW}. The form of
CDW order proposed generally differs from that discussed here (being more 
one-dimensional), and the models still do not explain the experimental 
observation that charge ordering arises at higher temperatures that spin 
ordering.  Here we discuss experimental evidence linking the phonon 
anomalies specifically to charged stripes.  It should be noted that there must
in general be a coupling to all modes: the charge modulation will act to scatter
all phonons; here we are speaking of a stronger coupling, with phonons localized
on the stripes.

It is by now clear that the LTT 
phase plays an important role in stabilizing a nearly static stripe 
phase\cite{Tran} in La$_{2-x-y}$R$_y$A$_x$CuO$_4$ (LACO:R) with $R$ a rare earth
(RE), typically $Nd$ or $Eu$, and $A$ = Sr, Ba, or Ca.  The reasons for this 
stabilization are less clear.  One explanation\cite{Tran} is a simple pinning 
effect: the stripes `prefer' to run parallel to the Cu-O-Cu bonds, and the LTT 
distortions lock them into an ordered array.  However, it is not necessarily 
true that LTT order enhances stripe pinning: in Eu substituted samples, the Cu 
spins appear to be even more dynamic in the LTT phase\cite{Kat3}.  Moreover,
such a picture completely fails to explain the common occurence of {\it 
simultaneous} fluctuating stripe order and fluctuating LTT order.  States with
fluctuating local order are relatively uncommon, and for two dynamic phases to
appear simultaneously and to interact strongly, but to have completely 
independent origins stretches credulity. 
It is much more likely that the connection is fundamental:
that the structural and stripe orders are two aspects of a single phenomenon. 
Indeed, this must be the case: neutrons are insensitive to individual electrons,
so the neutron diffraction evidence for long range stripe order\cite{Tran} is
really detecting an accompanying structural order.  Perhaps part of the problem
is a limited vocabulary for dealing with fluctuating local structural order: 
the {\it average macroscopic} structure will generally appear to be LTO or LTT
or some intermediate structure; only if the domains are of sufficiently large
scale would one expect to see a clear mixture of two phases.

Hence, this Appendix deals specifically with evidence that the structural order
is also short range, and evolves with doping in parallel with the electronic
stripes.  The cumulative evidence supports a picture of coexisting structural
and electronic stripe structure: whenever stripes are present, not only are the 
electronic properties inhomogeneous (with insulating magnetic and conducting 
charged stripes), but also the lattice distortions, with local LTT distortions 
on the charged stripes, while low-temperature orthorhombic (LTO) distortions 
are favored by the magnetic stripes.  This strongly supports a model of the 
stripe phase as a form of nanoscale phase separation, with the charged-stripe 
phase stabilized by an LTT-like phase\cite{RM8b,RM8cd}.  

\subsection{Coupling of Charged Stripes to local LTT Order}

\subsubsection{LTT-LTO Competition}

Uniform, macroscopic LTT order is rare in the La$_{2-x}$Sr$_x$CuO$_4$ (LSCO) 
family of cuprates.  There is a narrow regime of LTT phase near optimal doping 
in La$_{2-x}$Ba$_x$CuO$_4$ (LBCO), while long-range LTT order can only be 
induced in LSCO by replacing some La with a RE, typically Nd or Eu. 
However, {\it short range} LTT order appears to be more pervasive.

EXAFS measurements by Bianconi and coworkers found evidence 
for anomalous long Cu-O bonds in doped Ba$_2$Sr$_2$CaCu$_2$O$_8$ 
(BSCCO)\cite{Bia1} and LSCO\cite{Bia2}, which were 
interpreted in terms of stripes with alternating LTO-LTT orders.
A number of experiments\cite{LTT1,LTT4,LTT2,LTT3,Boz} have explored the 
possibility\cite{RM8cd} that the high-temperature tetragonal (HTT) and 
LTO phases can be `dynamic LTT' phases (i.e., with local LTT order).
Billinge, et al.\cite{LTT1} analyzed neutron diffraction pair distribution
functions (PDF's) of LBCO, and found that in the LTO phase the local order
has LTT symmetry, crossing over to average LTO structure on a $~10\AA$ length 
scale.  Moreover, there is no change in the local structure when warming into 
the HTT phase.  EXAFS results are consistent\cite{LTT4}.  In LSCO, Bo\v zin, et 
al.\cite{LTT3,Boz} find a crossover from local LTO order very near half filling, 
to a clear mixing of local LTO and LTT order in the doped materials.  The 
results are influenced by large local distortions near the dopants, particularly
in LBCO associated with the large difference in ionic size between Ba$^{+2}$ 
and La$^{+3}$,\cite{LTT4} but also near the dopant Sr sites\cite{LTT2}.  It has
been suggested\cite{PHS} that the distortions are characteristic of the 
hole-doped CuO$_2$ planes, but with enhanced distortion near Sr due to partial 
hole localization.  These doped materials display considerable tilt and 
bond-length disorder, interpretable in terms of stripes.  

A crossover from local LTO to local LTT order with increased doping is an 
intriguing possibility, particularly in light of the crossover in {\it stripe 
orientation} from diagonal at low doping\cite{diag}, which could be `pinned' by 
LTO order, to horizontal near optimal doping.  (It has been suggested that the
LTT distortions orient the stripes along the Cu-O bonds\cite{KSW}.)  The 
connection between the LTO phase and magnetic stripes may be related to 
strong magnetoelastic coupling: in the N\'eel phase, the spin flop transition is
strongly suppressed by LTT order\cite{Kat3}, while in the diagonal stripe phase 
the cluster spin glass transition can be seen in anelastic relaxation 
measurements\cite{Cord}.  The Mott insulator Sr$_2$CuO$_2$Cl$_2$ (SCOC) is found
by x-rays to be tetragonal, but below the N\'eel temperature the infrared
absorption peak associated with the Cu-O bending mode phonon is found to split
into two components\cite{Zib}.

\subsubsection{LTT and Stripes}

Recent studies have added considerable evidence for direct phonon coupling to 
stripes.  First, of course, is the direct neutron diffraction evidence for
charge order\cite{Tran,Mook}: what is seen by neutrons is not the electronic 
charges themselves, but an accompanying lattice distortion, suggesting that the
distortion is a fundamental part of the stripes.  These anomalies, and related 
anomalies of the half-breathing mode\cite{McQ1} need not imply that stripes 
are stabilized by electron-phonon coupling\cite{RM3}: it could simply be that 
there is stronger electron-phonon coupling on hole-doped stripes.  On the other 
hand, Bianconi, et al.\cite{Bian} report EXAFS evidence for a quantum critical 
point (QCP) -- a local splitting of the Cu-O bond length which can be tuned
to zero by adjusting the chemical microstrain.  It is found that the highest 
critical temperatures are associated with this optimal degree of
`microstrain'\cite{Bian}.  The associated bondlength splittings are much
larger than can be accounted for by the doping dependence of the average lattice
constant, and hence point to a structural QCP distinct from stripe formation.
Finally, there is the 
finding\cite{Boz} that local LTT-LTO type structural disorder is characteristic 
of the stripe regime, and the structural disorder greatly reduces near $x=0.25$,
close to the point where the stripe phase terminates\cite{Tal1,OSP}.  This 
near coincidence of stripe disorder and structural disorder provides very strong
evidence that the two kinds of stripe are associated with two kinds of 
structural order: that the LTO phase is connected with magnetic stripes, at 
half filling, and the LTT phase with charged stripes, near optimal doping.  The
data {\it cannot} be understood in terms of preexisting stripes accidentally
pinned by a uniform LTT order.

\begin{figure}
\leavevmode
   \epsfxsize=0.33\textwidth\epsfbox{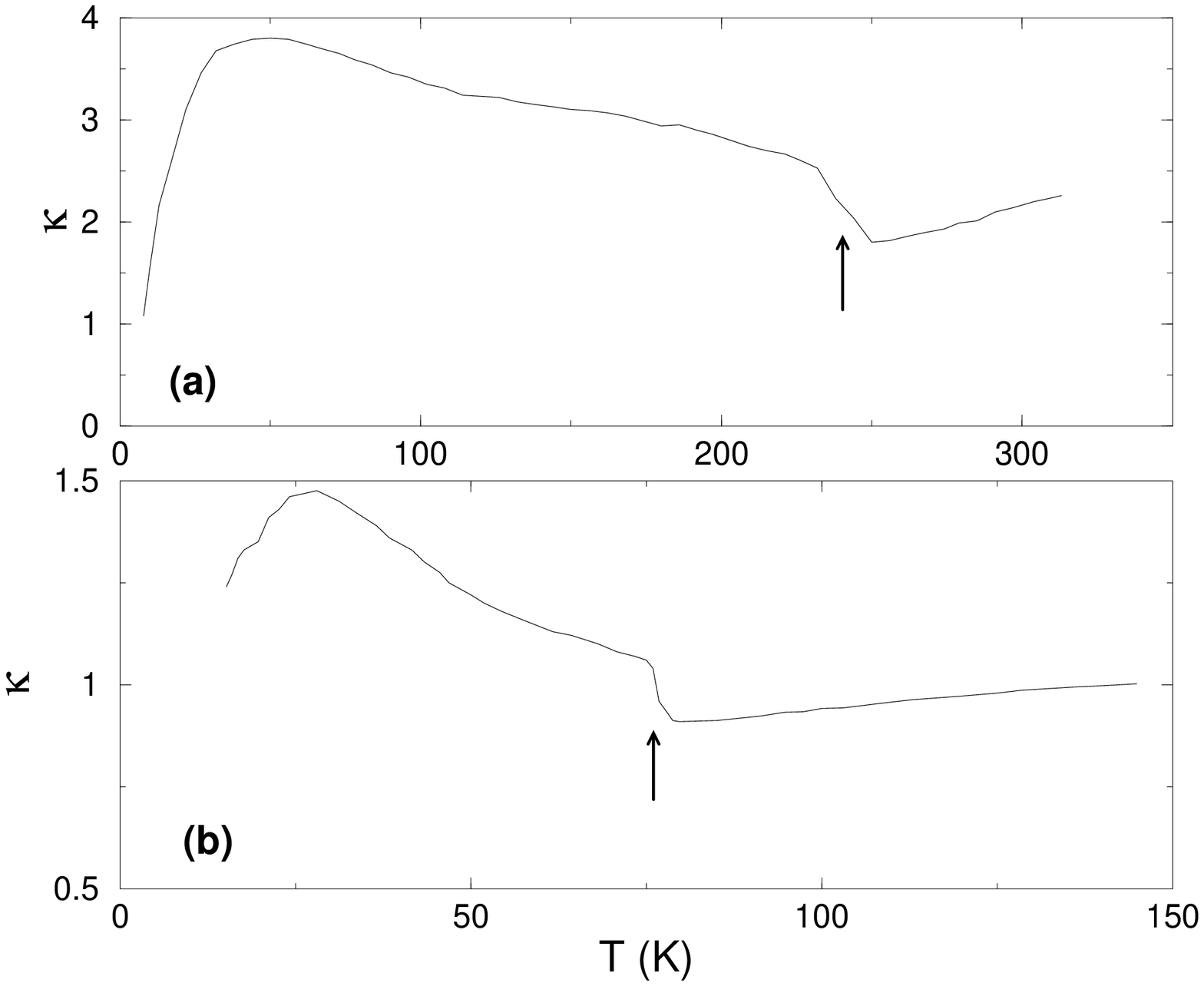}
\vskip0.5cm 
\caption{Thermal conductivity of (a) La$_{0.525}$\-Pr$_{0.1}$\-Ca$_{0.375}$MnO$
_3$
[\protect\onlinecite{Kim}], and (b) La$_{1.28}$Nd$_{0.60}$Sr$_{0.12}$CuO$_4$
[\protect\onlinecite{Bab}].}
\label{fig:32}
\end{figure}

Thermal conductivity measurements suggest a very close connection between 
stripe order and the LTT phase.  Thermal conductivity $\kappa$ has proven to be 
an important probe of the charge ordering transition in nickelates and 
manganites\cite{CHess}.  Typically, $\kappa$ is suppressed above the transition
due to collisions between phonons and fluctuating 
stripes.  Baberski, et al.\cite{Bab} find a very similar suppression of $\kappa$
in the cuprates, but {\it always at the LTT transition}.  Figure~\ref{fig:32} 
compares the thermal conductivity near the LTT transition in 
LSCO:Nd[\onlinecite{Bab}] (b) with that of the charge ordering transition in a 
manganite\cite{Kim} (a), suggesting an intimate connection between the LTT and 
charge ordering transitions.  In the absence of long range LTT order, $\kappa$ 
is suppressed over the full temperature range -- this is found both for pure 
LSCO and for the RE substituted materials, for $x>0.17$.
The thermal conductivity suppression persists down to the lowest doping studied,
$x=0.05$, and the jump at $T_{LTT}$ actually increases with decreasing $x$.  
However, suppression is absent in the undoped materials.  It would be most
interesting to see if the suppression terminates near $x=0.05$, where there is a
superconductor-insulator transition and the stripes cross over from horizontal 
to diagonal.

\begin{figure}
\leavevmode
   \epsfxsize=0.33\textwidth\epsfbox{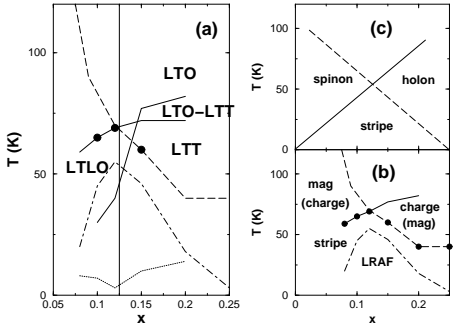}
\vskip0.5cm 
\caption{(a) Phase diagram of La$_{1.6-x}$Nd$_{0.40}$Sr$_x$CuO$_4$
[\protect\onlinecite{Ich}].  Solid lines = structural transitions, with various
phases labelled (LTO-LTT = mixed phase, LTLO = low temperature less orthorhombic
phase).  Long dashed lines = NQR knockout line [\protect\onlinecite{Ima}], 
indicating onset of slow stripe fluctuations; 
circles = charge order; dot-dashed line = long-range (incommensurate) 
antiferromagnetic (LRAF) order; dotted line = superconductivity. (b) 
Interpretation, in terms of dominant and subdominant (in parentheses) 
fluctuations. (c) Spin-charge separation phase diagram
[\protect\onlinecite{hosp}]: below the dashed line spinons are paired, below
the solid line holons are bose condensed.}
\label{fig:32b}
\end{figure}

The close connection between the LTT phase and charged stripes is further
illustrated by a recent study of the rich phase diagram found\cite{Ich,Ima} in
La$_{1.6-x}$Nd$_{0.40}$Sr$_x$CuO$_4$, Fig.~\ref{fig:32b}a.  This complex 
behavior can be simply
understood (Fig.~\ref{fig:32b}b) if it is kept in mind that the structural 
phases are macroscopic averages, while local structure can be more complicated.
We assume that the LTO phase is associated both with the high-temperature
electronically disordered phase, and with the magnetically ordered phase at $x=
0$, while the LTT phase is characteristic of hole-doped stripes.  As temperature
is decreased, long-range stripe order arises in two steps: first the charges 
order (filled circles) then the spins (dot-dashed line).  The other curves in
the diagram are associated with fluctuating stripe order.  Thus, the NQR 
`knockout' line may indicate the onset of stripe fluctuations\cite{Ima}; at 
nearly the same phase boundary XANES\cite{Sai} and PDF\cite{Boz}  measurements 
find the onset of local Cu-O bondlength variations suggestive of local LTT-type 
tilts.   Near 1/8 doping (vertical line) there is a crossover in the
stripe fluctuations: at low doping there are narrow charged stripes fluctuating
in a magnetic (LTO) background; at higher doping narrow magnetic stripes
fluctuate in a charged (LTT) background.  Long range charge order (filled 
circles) coincides with the {\it lower} of the LTT(-like) transition or the
NQR knockout line.

Magnetic field studies provide additional evidence.  The Hall effect\cite{Uch} 
in strongly Nd substituted LSCO shows a crossover from one-dimensional ($R_H
\rightarrow 0$ as $T\rightarrow 0$) to two-dimensional behavior at $x=0.12$; for
$x< 0.12$, the $R_H$ anomaly turns on with stripe order -- exactly at the 
LTT-like phase transition.  Further, it is known that an in-plane magnetic field
has a strong effect on magnetoresistance, including hysteretic effects which can
be interpreted as field-induced rotation of the stripes\cite{Ando1}.
Remarkably, an in-plane field also has a strong effect on the LTT-LTO
transition temperature\cite{XON}, stabilizing the LTT phase at higher $T$.  This
can be qualitatively understood, if the LTT phase and charge stripes are 
intimately related: the fluctuating (LTT) stripes are present in the LTO phase, 
but the strong fluctuations produce an average LTO structural order; the field, 
by aligning the stripes, reduces transverse fluctuations and reveals the 
underlying LTT order.

In summary, the data suggest the following scenario.
The magnetic stripes are associated with a local LTO-type tilt distortion, the
charged stripes with an LTT tilting.  At half filling, there is a simple
long-range LTO order, which may be stabilized by magnetoelastic coupling
which aligns the spins with the LTO tilt direction.  
The LTT tilting, stabilized by coupling to the VHS, is fluctuating unless 
pinned by ionic disorder off of the CuO$_2$ planes.  When the 
LTT phase is pinned, the stripes develop long-range charge (and ultimately spin)
order.  If the stripe pinning is strong enough, the superconducting order can be
suppressed, predominantly due to the magnetic order\cite{Ich}.  Near enough to 
half filling, the LTO tilts prevail, the stripes rotate by 45$^o$, and 
superconductivity is destroyed.

Thus there is a three-way competition, between magnetic, structural, and
superconducting order.  Such a model could explain why superconductivity is not
found in the nickelates and manganites, where electron-phonon coupling is
stronger.  The competition between superconductivity and LTT order takes place
predominantly {\it on the charged stripes}\cite{OSP,Surv}, and is strongly
influenced by interlayer strains, which can be varied by rare earth 
substitution.  B\"uchner, et al.\cite{Buch} find a critical tilt angle $\Phi_c$,
such that when the average LTT tilt exceeds $\Phi_c$, `bulk' superconductivity
is absent (the Meissner fraction is greatly reduced), while for tilts less than
$\Phi_c$ superconductivity is hardly affected by the presence of LTT order.
Experimentally, it is often found that when LTT fluctuations are weak (e.g., in
some compositions of LSCO and oxygen-doped La$_2$CuO$_4$\cite{LTO}), 
magnetic order develops exactly at the superconducting transition.  Julien, et
al.\cite{Jul} have suggested that magnetic fluctuations are present at much 
higher temperatures, and that the sudden slowing of the fluctuations at $T_c$ is
coincidental.  We suggest a slightly different possibility: that when LTT 
pinning is particularly weak, it takes superconducting order to provide the 
stripe phase with structural rigidity, and that it is the stiffening of the
charge stripes which is reflected in the slowing of the  magnetic fluctuations. 
This provides one more demonstration that the charged stripes can be either 
superconducting or LTT ordered.

\subsection{Flux Phase on Stripes}

Early studies of the flux phase generally concentrated on the competition
between the flux and antiferromagnetic phases at half filling.  The present 
analysis strongly suggests that the flux phase would be present instead 
in the hole doped region -- that is, on the charged stripes.  This follows both
because instability is enhanced near a VHS, and because the correlated hopping
scales with $x$; moreover, in a stripe picture, features near the Fermi level 
are generally associated with the charged stripes\cite{OSP}.  This conclusion
is consistent with the experimental observation\cite{Moofl} that the possible
flux related magnetization is stronger in YBCO$_{6+y}$ for $y$=0.6 than for
$y$=0.35 (see also Chakravarty, et al.\cite{CKN}).  
It also provides a natural resolution 
of the Lee-Wen paradox\cite{LeW} discussed by Orenstein and Millis\cite{OM}.
Lee and Wen showed that they could explain the Uemura relation for underdoped
cuprates, as long as the flux phase dispersion is independent of doping.  The
paradox is that many strong-coupling models expect the dispersion to renormalize
to zero near the Mott insulator at half filling.  In a stripe picture, this
renormalization is taken as indicating that the {\it fraction of material}
associated with charged stripes renormalizes to zero at half filling, whereas
{\it the dispersion on a single stripe} is less sensitive to doping\cite{OSP}.

At very low doping, flux-stripes will tend to break up into flux-polarons, 
confined to a single plaquette.  These flux-polarons bear a close resemblance to
the skyrmions introduced by Gooding\cite{RGood}.  Indeed, his electronic states 
are just linear combinations of the $E_u$-symmetry plaquette states of Eq. 
\ref{eq:11z}.  The difference is that his states are localized around a Sr 
impurity, while ours are the equilibrium conducting state on the hole-doped 
stripes.  Clearly, at low temperatures there will be a tendency for charged 
stripes and polarons to be pinned on the Sr, greatly enhancing the similarity.  
However, the present model more naturally explains stripes at higher doping 
levels, and the uniform charged phase near $x_0$.  It should be noted that 
Haskel, et al.\cite{LTT2} find enhanced local structural distortions near Sr 
impurities, associated with hole localization\cite{PHS}.

If the flux phase is on the charged stripes, then hole doping will shift the 
Fermi level away from the conical points, producing not a node but a hole 
pocket, as in Fig.~\ref{fig:3}).  It is possible that the deviations from d-wave
symmetry found in the underdoped regime\cite{Mes1} are associated with this hole
pocket.  The apparent d-wave gap might then
be a {\it localization} gap at the hole pocket Fermi surface, leading to the
localization effects observed in resistivity\cite{Boeb,Quit}.

\subsection{Spinons and Holons and Stripes}

\begin{figure}
\leavevmode
   \epsfxsize=0.33\textwidth\epsfbox{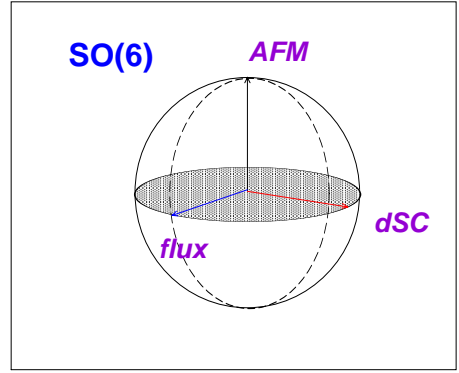}
\vskip0.5cm 
\caption{Six dimensional order parameter in SO(6) theory 
[\protect\onlinecite{MarV}], with three components of magnetic order (AFM), two
of d-wave superconductivity (dSC), and one of flux phase (flux).  In a stripe
picture, the shaded plane would correspond to the charged stripes.}
\label{fig:39}
\end{figure}

Overall, the idea that stripes involve competition between
antiferromagnetism, d-wave superconductivity, and the flux phase is very 
attractive, and arises naturally from an SO(6) symmetry (a subgroup of the
SO(8) group\cite{MarV}),
in which all three states comprise a single 6-dimensional order parameter,
Fig.~\ref{fig:39}.  In fact, in the repulsive Hubbard model at half filling, 
when $t'=0$, there is an extra electron-hole symmetry, leading to a degeneracy 
between the flux phase and d-wave superconductivity (dSC)\cite{And}.  This is 
equivalent to the degeneracy between a CDW and s-wave superconductivity (sSC) in
the attractive Hubbard model, since the flux phase is a form of d-wave CDW.  In 
fact, the $\tau$ operator of Eq.~\ref{eq:12b} transforms the OAF and dSC
operators into the CDW and sSC operators\cite{MarV}.
This degeneracy is the basis for the SU(2) model of superconductivity\cite{LeW}.
The SO(6) model thus encompasses the SO(5) and SU(2) models, Fig.~\ref{fig:39}.

There is a close similarity between the experimental doping dependence of LSCO, 
Fig.~\ref{fig:32b}a,b and the spinon-holon phase diagram\cite{hosp,SU2}, 
Fig.~\ref{fig:32b}c, with the spinon-condensed phase corresponding to the
magnetic stripe dominated regime and the holon-condensed phase corresponding 
to the charged stripe dominated regime, as indicated by the experimental LTT 
phase, Fig.~\ref{fig:32b}a!  Strikingly, the holon condensed regime is where the
holes have a flux-phase Fermi surface\cite{SU2}.  Indeed, it would seem that 
spin-charge separation
can be interpreted as a theory of stripe phase formation.  Thus, the spinon 
pairing involves the exclusion of spin deviations from an antiferromagnetic 
domain onto an antiphase domain wall, while holon condensation involves the 
ordering which stabilizes the charged domains (the flux phase).  The spin gap 
would then correspond to a quantum confinement energy\cite{OSP}.  When both 
spinons and holons are condensed together, well-defined stripe order results.  
As early as 1987, Anderson\cite{And2} noted that spinon and holon condensation
should couple to structural distortions.  Superconducting order seems to be more
characteristic of the holon condensed domain (charged stripes) than of the 
stripe phase regime.

\end{document}